\documentclass[prb,aps,twocolumn,english,amsmath,amssymb,floatfix,superscriptaddress,longbibliography]{revtex4-2}

\usepackage{amsbsy}
\usepackage{amsfonts}
\usepackage{babel}
\usepackage{bm}  
\usepackage{color}
\usepackage{dcolumn}
\usepackage{enumerate}
\usepackage{epsf}
\usepackage{esint}
\usepackage{epsfig,psfrag}
\usepackage{float}
\usepackage{graphicx}
\usepackage{latexsym}
\usepackage{mathbbol}
\usepackage[usenames,dvipsnames]{xcolor}
\usepackage[mathscr]{eucal}
\usepackage{wrapfig}

\definecolor{goodgreen}{rgb}{0.1,0.5,0}
\definecolor{goodred}{rgb}{0.7,0,0}
\usepackage[colorlinks,urlcolor=goodgreen,citecolor=blue,linkcolor=goodred]{hyperref}

\usepackage{cancel}

\newcommand{\tr}{\mathrm{tr}}

\newcommand{\sign}{\text{sign}}
\newcommand{\comment}[1]{}

\usepackage[normalem]{ulem}

\begin{document}

\title{Chiral spin channels in curved graphene \emph{pn}~junctions}

\author{Dario Bercioux}
\email{dario.bercioux@dipc.org}
\affiliation{Donostia International Physics Center (DIPC), Manuel de Lardizbal 4, E-20018 San Sebasti\'an, Spain}
\affiliation{IKERBASQUE, Basque Foundation for Science, Euskadi Plaza, 5, 48009 Bilbao, Spain}

\author{Diego Frustaglia}
\email{frustaglia@us.es}
\affiliation{Departamento de F\'isica Aplicada II, Universidad de Sevilla, E-41012 Sevilla, Spain}

\author{Alessandro De Martino}
\email{ademarti@city.ac.uk}
\affiliation{Department of Mathematics, City, University of London, 
London EC1V 0HB, United Kingdom}

\date{\today}
\begin{abstract}
We show that the chiral modes in circular graphene $pn$~junctions provide an advantage 
for spin manipulation via spin-orbit coupling compared to semiconductor platforms. 
We derive the effective  Hamiltonian for the spin dynamics of the junction's zero modes 
and calculate their quantum phases. We find a sweet spot in parameter space 
where the spin is fully in-plane and radially polarized for a given junction polarity. 
This represents a shortcut to singular spin configurations that would otherwise require 
spin-orbit coupling strengths beyond experimental reach.
\end{abstract}

\maketitle

\section{Introduction}

Graphene has attracted exceptional
interest as a quantum material with Dirac cones at the Fermi energy 
and other unique electronic properties~\cite{Novoselov:2004,Novoselov:2005,CastroNeto:2009}. 
One appealing  feature is the possibility of tuning electrostatically 
the   charge carriers' polarity in $pn$~junctions  of
linear~\cite{Abanin:2007,Williams:2007dr,Williams:2011da,Williams:2011kb,Ozyilmaz:2007fj,Rickhaus:2015cpa} 
and circular shape~\cite{Freitag_2016,Zhao_2015,RodriguezNieva_2016,Ghahari_2017,Jiang_2017,
Gutierrez2018,Brun_2019,Ren2021,Brun_2022}. 
The latter have been created by different means, such as the tip potential of 
a scanning tunneling microscope~\cite{Freitag_2016,Jiang_2017,Brun_2019,Brun_2022} 
or by placing impurities in the substrate~\cite{Zhao_2015,Ghahari_2017}. In both approaches, 
experiments have shown that  it is possible to single out and steer individual electronic eigenstates. 
Importantly,  $pn$~junctions are essential building blocks for graphene-based electron-optical elements 
and edge-state interferometers~\cite{Mrenca_Kolasinska_2016,Jiang_2017,Jo_2021,Waintal_2022} 
also exploiting the so-called snake states~\cite{Rickhaus:2015cpa,Cohnitz_2016,makk2018}.

The electronic spin degree of freedom is usually neglected in the study of graphene
$pn$~junctions because of the weak atomic spin-orbit coupling (SOC) of 
carbon~\cite{HuertasHernando_2006,Boettger2007,Gmitra2009,Bercioux_2015}. 
However, theoretical predictions followed by experimental realizations 
proved that strong SOCs  can be induced, e.g., by proximity with transition metal dichalcogenide (TMD)
substrates~\cite{Gmitra:2015kg,Gmitra:2016fk,Wang_2015,Wang_2016,Wakamura2020,Wakamura2021,Wang_2021,Naimer_2021,Naimer_2023,Tiwari_2022}.
These advances open the exciting possibility of including the spin functionality in graphene-based 
electron optics, with the further benefit that the versatility of $pn$~junctions allows 
for the design of curved waveguides for spin and charge carriers. This is particularly 
interesting in view of the intense current theoretical and experimental research 
activity on the spin dynamics triggered by SOC in curved geometries~\cite{Das_2019,Frustaglia_2020,Streubel_2021,Gentile_2022}.  The effects of SOC in graphene have been also investigated in other geometries~\cite{Zarea_2009,De_Martino_2011,Lenz_2011,Lenz_2013}.

In this article, we investigate circular $pn$~junctions in the presence of 
(i) a perpendicular magnetic field, coupled to the electronic charge 
(developing Landau levels in the quantum Hall regime) and spin (through Zeeman coupling), 
and (ii) proximity-induced SOCs of different types. 
We provide the exact solution of graphene's Dirac equation for this system and formulate an effective one-dimensional (1D) model for 
the spin and angular dynamics of the states localized at the $pn$~interface. 
This resembles the model for semiconductor rings subject
to Rashba SOC (RSOC)~\cite{Frustaglia2004}, 
with a meaningful difference: the chiral nature of the propagating modes. 
We identify a remarkable sweet spot in the parameter space, 
where the spin eigenstates align locally with the effective magnetic field produced by the SOC.
This point coincides with the Rabi condition for electronic spin resonance in a magnetic field
and represents a shortcut to adiabatic spin dynamics unavailable in its semiconductor equivalent. 
We confirm this result within the original full model and propose a set-up to identify 
this sweet spot via spin interferometry, opening a promising  route to spin state manipulation in graphene. 

The article is organized in the following way: In Sec.~\ref{model}, we introduce the model system. In Sec.~\ref{lwm}, we present a low-energy model for the system under investigation, where we show the presence of the sweet spot in the parameter space. In Sec.~\ref{exppro}, we provide a proposal for an interferometric experiment to detect the presence of this sweet spot. We discuss in Sec.~\ref{discussion} the interpretation of the experimental proposal and its range of validity. Finally, in Sec.~\ref{conclusions}, we provide our conclusions. All the technical details are presented in the Supplemental Material~(SM)~\cite{supplemental}.

\section{Model}\label{model}
The low-energy model for graphene with proximity-induced SOCs reads
%
%
\begin{equation}
    \mathcal{H} = {\mathcal H}_0 + \mathcal{H}_\text{spin},
\label{ham}
\end{equation}
%
%
where $\mathcal{H}_0$ is the Dirac Hamiltonian in a perpendicular magnetic field
%
%
\begin{align}
{\mathcal H}_0 &=v_\text{F} \left( \tau \sigma_x \Pi_x + \sigma_y \Pi_y \right)   
+ V,
\label{ham0}
\end{align}
%
%
with Fermi velocity $v_\text{F}$ and kinetic momentum ${\bm \Pi} = -i \hbar   \bm{\nabla}+ \frac{e}{c} \bm{A}$,
with $\bm{A}=\frac{B}{2}(-y,x)$ in the symmetric gauge. 
Here, $\tau=\pm 1$ denotes the valley index and 
$\bm{\sigma}=(\sigma_x,\sigma_y)$ are Pauli matrices in sublattice space~\cite{CastroNeto:2009}.
%
%
\begin{figure}[!t]
    \centering
    \includegraphics[width=.95\columnwidth]{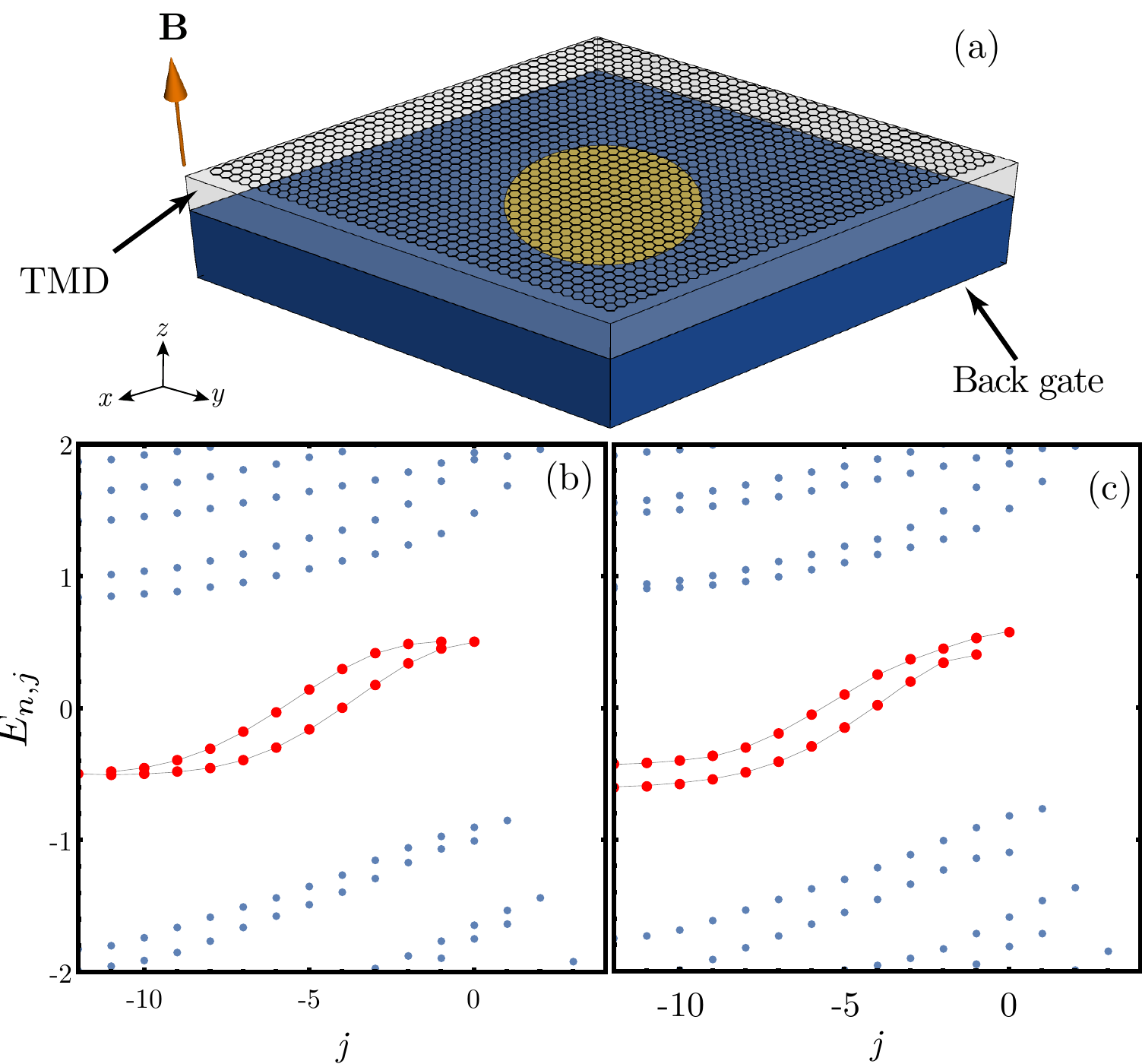}
    \caption{(a) Sketch of the system, with $p$ and $n$ regions
    drawn in yellow and blue. 
    (b) Energy spectrum versus angular momentum $j$ for $V_0=0.51$, $\xi_0=5.1$, 
    $\lambda_\text{R}=0.5$ and $\lambda_\text{Z}=\lambda_\text{KM}=0$. 
    (c) Same as in (b) but for $\lambda_\text{Z}=0.1$. 
    In (b) and (c), the red dots highlight the zeroth Landau levels.}
    \label{fig_one}
\end{figure}
%
%
The potential 
%
%
%
\begin{equation}
\label{potential}
V(r)= V_0 \, \sign(R-r),
\end{equation}
%
%
defines a circular $pn$~junction of radius $R$, with a $p$-doped region for $r<R$ (the ``dot"), 
and a $n$-doped region for $r>R$.
The system is sketched in Fig.~\ref{fig_one}(a).
The spin-dependent part ${\mathcal H}_\text{spin}= \mathcal{H}_\text{Z} + \mathcal{H}_\text{R} + \mathcal{H}_\text{KM} + \mathcal{H}_\text{VZ}$
includes both  Zeeman and SOCs terms~\cite{Gmitra:2015kg,Gmitra:2016fk,Frank_2020}: 
%
%
\begin{subequations}
\begin{align}
\mathcal{H}_\text{Z}  &=\lambda_\text{Z}s_z
, \\
\mathcal{H}_\text{R} & =  \frac{\lambda_\text{R}}{2} \left( \tau \sigma_x s_y - \sigma_y s_x \right),\\
\label{RSOC_Zeeman}
\mathcal{H}_\text{KM}  & =   \lambda_\text{KM} \tau \sigma_z s_z,\\
\mathcal{H}_\text{VZ} & =  \lambda_\text{VZ}   \tau s_z.
\end{align}
\end{subequations}
%
%
Here $\lambda_\text{Z}=\frac{g_s  \mu_B}{2}B$, and $\bm{s}=(s_x,s_y)$ denotes 
the Pauli matrices in spin space. The terms $\mathcal{H}_\text{R}$, $\mathcal{H}_\text{KM}$, 
and $\mathcal{H}_\text{VZ}$ are  the Rashba,  Kane-Mele, and 
valley-Zeeman SOC, respectively~\cite{HuertasHernando_2006,Kane_2005a,Gmitra2009}. 
Precise estimates for the SOCs depend on the specific heterostructure, 
e.g., the relative orientation between graphene and substrate~\cite{Naimer_2021,Naimer_2023}.
The RSOC and the VZSOC range from few hundredths of meV up to few meV, 
while the KMSOC is typically much smaller~\cite{Naimer_2021,Tiwari_2022}.
We are mainly concerned with the effects of the Zeeman and RSOC terms.
The valley-Zeeman term can be included by means of a valley-dependent shift of the Zeeman coupling 
and will be considered separately in the discussion section below. 
For $\lambda_\text{VZ}=0$, the valley degree of freedom just leads to a degeneracy factor, 
so we can focus on a single valley and set $\tau=+1$.
Throughout this paper, we measure 
lengths in units of magnetic length $\ell_B=\sqrt{\hbar c/e B}=25.65~\text{nm}/\sqrt{B[\text{T}]}$ 
and  energies in units of cyclotron energy $\hbar\omega_\text{c}=\hbar v_\text{F}/\ell_B\approx 26~\text{meV}\sqrt{B[\text{T}]}$, and assume a typical field $B\sim 1$~T~\cite{Tiwari_2022}.

In this model, the wave function is a four-component spinor 
$\Psi^\text{T}= \left( \Psi_{\text{A}\uparrow},\Psi_{\text{B}\uparrow},
\Psi_{\text{A}\downarrow},\Psi_{\text{B}\downarrow}\right)$.
The Hamiltonian ${\mathcal H}$ commutes with the total angular momentum 
$J =  L_z + \frac{1}{2}\left( \sigma_z +s_z\right)$,
with $L_z=-i\partial_\theta$ the orbital angular momentum, 
hence its eigenstates $\Psi_j (\mathbf{r})$, expressed in terms of confluent hypergeometric functions~\cite{Olver_2010,supplemental,DeMartino_2007,DeMartino_2010}, can be labelled by an integer $j\in \mathbb{Z}$. 
The spectrum is illustrated in Figs.~\ref{fig_one}(b) and~\ref{fig_one}(c). 
In particular, we find two ``zero-energy" Landau levels (LLs),
the ``top" (T) and ``bottom"  (B) zero modes,
highlighted in red in the figures. 
In the absence of SOCs, they have zero energy for $V_0=0$,
but develop a dispersion in $j$ for finite $V_0$~\cite{Cohnitz_2016,supplemental}.
Their energy at $j=0$ and at $j\ll -1$ 
approaches the value of the potential $V(r)$ inside and outside the 
dot, respectively, see Fig.~\ref{fig_one}(b).
In the presence of RSOC, the two modes acquire a spin splitting, similar to the case 
of a two-dimensional electron gas (2DEG)~\cite{Bercioux_2015, Bercioux_2019}. 
A finite Zeeman coupling produces an additional vertical splitting|see Fig.~\ref{fig_one}(c). 
We present in the SM~\cite{supplemental} the exact solution of the
model~\eqref{ham}, including a detailed analysis of the spin splitting as a function of~$\lambda_\text{R}$.

\section{Effective 1D model}\label{lwm}

In order to describe the low-energy physics 
around the Fermi energy (set at the charge neutrality point, $E_\text{F}=0$),
we introduce an effective 1D Hamiltonian for  
the zero modes localized at the $pn$~interface.
We follow an analogous derivation for a semiconductor ring with RSOC~\cite{Meijer_2002}, 
see the SM~\cite{supplemental} for details. 
We first perform a unitary transformation, $\mathcal{H}\rightarrow \tilde{\mathcal{H}} =U\mathcal{H}U^{-1}$, 
with $U=e^{i\frac{\sigma_z}{2}(\theta+\frac{\pi}{2})}e^{i\frac{s_z}{2}\theta}$.
In this rotating frame, we factorize the wave function 
as $\tilde{\Psi}=\tilde{\psi}_0(r)\tilde{\chi}(\theta)$, 
where $\tilde{\psi}_0(r)$ is the sublattice spinor 
for the (spin degenerate) zero mode of the radial part of $\tilde{\mathcal{H}}_0$, 
and $\tilde{\chi}(\theta)$ is a spinor in spin space, containing the angular dependence. 
The projection of $\tilde{\mathcal{H}}$ onto the zero mode $\tilde{\psi}_0(r)$ leads to 
the effective 1D Hamiltonian controlling the dynamics of $\tilde{\chi}(\theta)$:
%
%
\begin{equation}
\tilde{\mathcal{H}}_\text{eff}=
 \omega_0  (L_z +\Phi)  + ( \omega_\text{Z}  - \frac{\omega_0}{2})s_z - \omega_\text{R} s_x.
\label{effectiveHam}
\end{equation}
%
%
The frequencies in Eq.~\eqref{effectiveHam} are defined by
%
%
\begin{subequations}
\begin{align}
 \omega_0 & =  \left\langle \frac{\sigma_x}{r} \right\rangle_0, \\
 \omega_\text{Z}&=  \lambda_\text{Z}  + \lambda_\text{KM} \langle \sigma_z\rangle_0, \\
 \omega_\text{R} &= \frac{\lambda_\text{R}}{2} \langle \sigma_x \rangle_0,
 \label{omega}
\end{align}
\end{subequations}
%
%
where $\langle \dots \rangle_0$ denotes the (radial) expectation value 
in the state $\tilde{\psi}_0(r)$. (We note that $\sigma_x$ 
is the azimuthal component of the velocity operator in the rotating frame.)
The parameter $\Phi \approx \xi_0= B\pi R^2/\Phi_0$ 
denotes approximately the magnetic flux 
through the dot in units of the flux quantum $\Phi_0$.
Since $\tilde{\mathcal{H}}_\text{spin}$ is treated perturbatively, this projection
is justified as long as $\hbar \omega_c$ is much larger than the Zeeman and SOCs.
The Hamiltonian~\eqref{effectiveHam} describes a 1D spinful chiral mode propagating along the
curved $pn$~interface, with angular velocity controlled by the gate voltage difference across the junction.
Importantly, the polarity of the junction determines the 
signs of $\omega_0$ and $\omega_\text{R}$~\footnote{We note that the signs of $\omega_0$ and $\omega_\text{R}$ 
are independent of the valley index.}.
For $V_0>0$ 
both are positive.
Inverting the polarity, $V_0\rightarrow -V_0$, reverses the propagation direction, changing both signs. 
This feature has crucial implications for the experimental setup discussed below.


Diagonalizing $\tilde{\mathcal{H}}_\text{eff}$, we obtain the eigenvalues
%
%
\begin{equation}
\label{Em-ring}
E_{m,\pm} =  \omega_0  (m+\Phi)  \pm  \sqrt{(\omega_\text{Z}-\frac{\omega_0}{2})^2+\omega^2_\text{R}} ,
\end{equation}
%
%
where $m\in\mathbb{Z}$ under periodic boundary conditions.  
This formula predicts a linear dependence of the energy on $m$, which 
we observe in the exact solution close to zero energy, and provides  
an approximate analytical expression for the slope of the  dispersion.  
%
%
\begin{figure}[!t]
    \centering
    \includegraphics[width=\columnwidth]{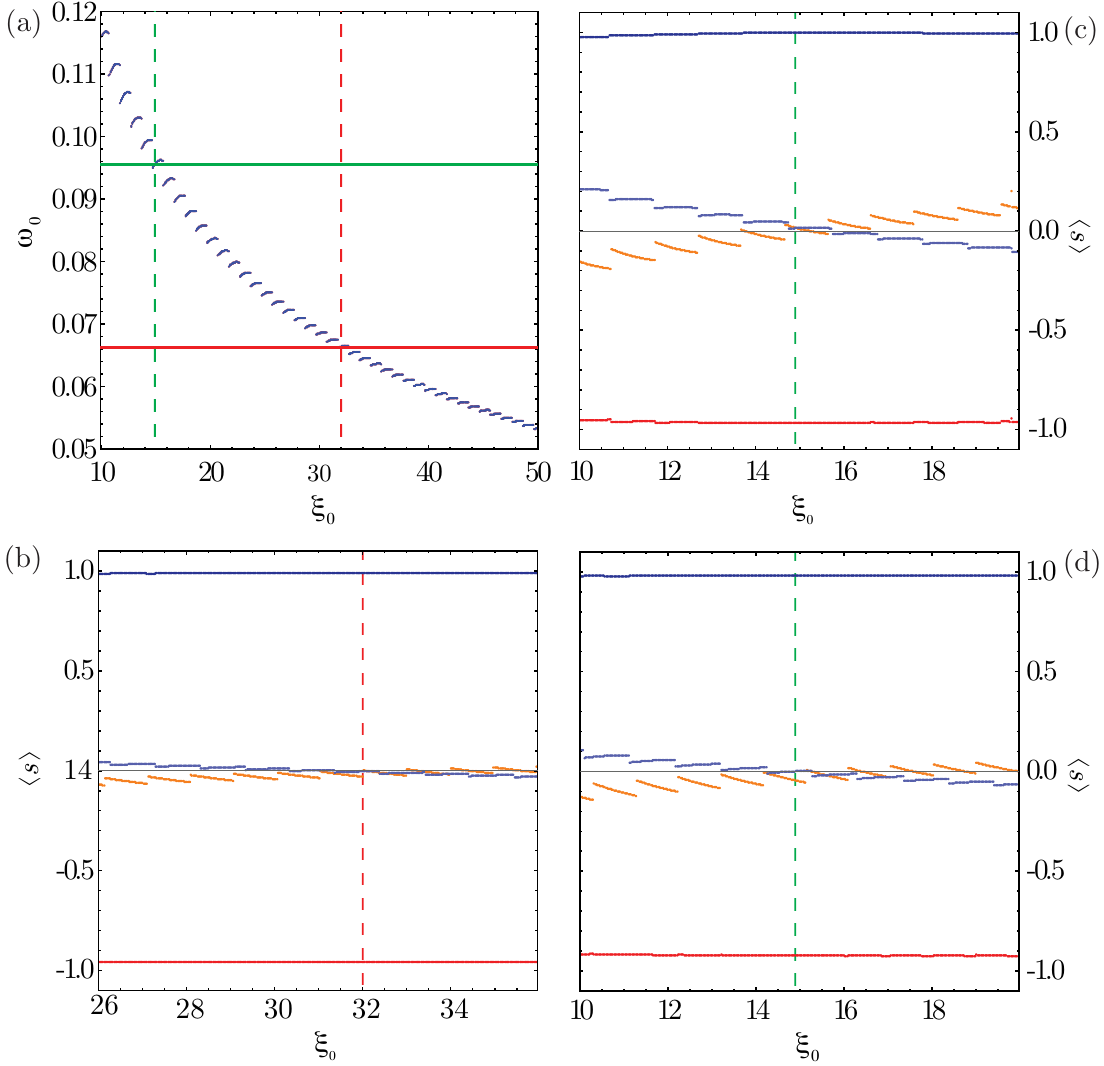}
    \caption{Sweet spot identification in the full model.
    (a) The angular frequency
    $\omega_0$ as a function of $\xi_0=R^2/2\ell^2_B$, 
    at a fixed magnetic field.    
    The green and red horizontal lines describe two representative values of $2\omega_\text{Z}$,
    and the vertical dashed lines the corresponding values of $\xi_0$  
    at which the resonance condition $2\omega_\text{Z}=\omega_0$ is realized. 
    (b)-(d) The exact expectation values of the radial and perpendicular
    spin components in the top and bottom modes as a function of 
    $\xi_0$, for $\lambda_\text{Z}=0.047$ and  $\lambda_\text{R}=0.2$ in (b) 
    and for $\lambda_\text{Z}=0.033$ and  $\lambda_\text{R}=0.2,0.3$ in (c) and (d) respectively. 
    In (b)-(d),
    the top curve shows $\langle s_r\rangle_\text{B}$, the bottom one $\langle s_r\rangle_\text{T}$, 
    and the two central ones $\langle s_z\rangle_\text{T}$ and $\langle s_z\rangle_\text{B}$. 
    In all the panels, $V_0=0.51$.\label{fig_two}}
\end{figure}
%
%
The corresponding eigenstates are
%
%
\begin{subequations}\label{eigenstates}
\begin{align}
\tilde{\chi}_{m,+} &=  \frac{e^{im\theta}}{\sqrt{2\pi}}
\begin{pmatrix}
\cos \frac{\gamma}{2} \\
-\sin \frac{\gamma}{2}
\end{pmatrix},\, \\
 \chi_{m,-} &=   \frac{e^{im\theta}}{\sqrt{2\pi}}
\begin{pmatrix}
\sin \frac{\gamma}{2} \\
\cos \frac{\gamma}{2}
\end{pmatrix},
\end{align}
\end{subequations}
%
%
where 
%
%
\begin{equation}
\label{gamma_angle}
e^{i\gamma}= \frac{\omega_\text{Z}-\frac{\omega_0}{2}+i\omega_\text{R}}{\sqrt{ (\omega_\text{Z}-\frac{\omega_0}{2})^2+\omega^2_\text{R} }}.
\end{equation}
%
%

We find a sweet spot for $\omega_{\rm Z}=\frac{\omega_0}{2}$ ($\gamma=\frac{\pi}{2}$), 
where the spin eigenstates~\eqref{eigenstates} point along the 
radial direction in the $xy$ plane for any value of $\omega_{\rm R}$. 
This situation is remarkable. It recalls the Rabi condition for spin resonance in the rotating 
wave approximation (RWA), with the difference that there is no Bloch-Siegert shift~\cite{Bloch1940} 
as a function of the driving amplitude (represented by $\omega_{\rm R}$): here, the RWA is exact.
Notice that an inversion of the junction polarity, changing the chirality of the propagating spin
channels ($\omega_0 \rightarrow -\omega_0$), would take the system off-resonance. 
This is in sharp contrast to the case of semiconductor-based Rashba rings~\cite{Frustaglia2004,Nagasawa2012}, 
where counter-propagating channels coexist, and a full in-plane alignment of the spinors 
is only achieved in the adiabatic limit of very large RSOC ($\omega_{\rm R} \gg \omega_0$)~\cite{Frustaglia2004}. 

The resonance condition, exact in the projected model~\eqref{effectiveHam}, 
holds with excellent accuracy also in the full model~\eqref{ham}.
This is shown in Fig.~\ref{fig_two}, where for simplicity we set $\lambda_\text{KM}=0$.
Here, we define the angular frequency $\omega_0$ as the expectation value
$\langle \sigma_x/r \rangle_{\lambda_\text{R}=0}$ on the $j$-state closest to zero energy. 
From Fig.~\ref{fig_two}(a), we can see that $\omega_0$ decreases 
as a function of the radius $R$  
and presents a staircase behavior due to the discreteness of $j$. 
In Figs.~\ref{fig_two}(b)-(d), we show the expectation values of the perpendicular 
and radial components of the spin, $s_z$ and $s_r$, in the top and bottom $j$-states closest 
to zero energy for different sets of parameters. 
We observe that at the value of $\xi_0$ where the resonance condition 
$\omega_{\rm Z}=\frac{\omega_0}{2}$ is realized, 
$\langle s_z\rangle$ is almost zero, whereas $\langle s_r\rangle$  is close to $1$.
The results in Fig.~\ref{fig_two} show an excellent agreement 
between the prediction of the projected model and the full solution. 
In particular, they confirm that the resonance condition is independent of the RSOC. 
The small discrepancies are due to the coupling of the zero modes 
to the higher LLs via the RSOC, neglected in the projected model.  
We present additional results, 
including the effect of~$\lambda_\text{KM}$, in the SM~\cite{supplemental}.

\section{Experimental proposal}\label{exppro}
We propose two setups based on linear 
and circular $pn$~junctions to implement interferometric circuits for spin carriers. 
Thanks to the chiral nature of the propagating channels, 
we find that, depending on the junction polarity,  
the interferometers respond differently to the Zeeman coupling $\omega_{\rm Z}$ (assuming $\lambda_{\rm KM}=0$ for simplicity), 
making possible a unique geometric characterization of the propagating spin states. 

\begin{figure}[!t]
    \centering
    \includegraphics[width=\columnwidth]{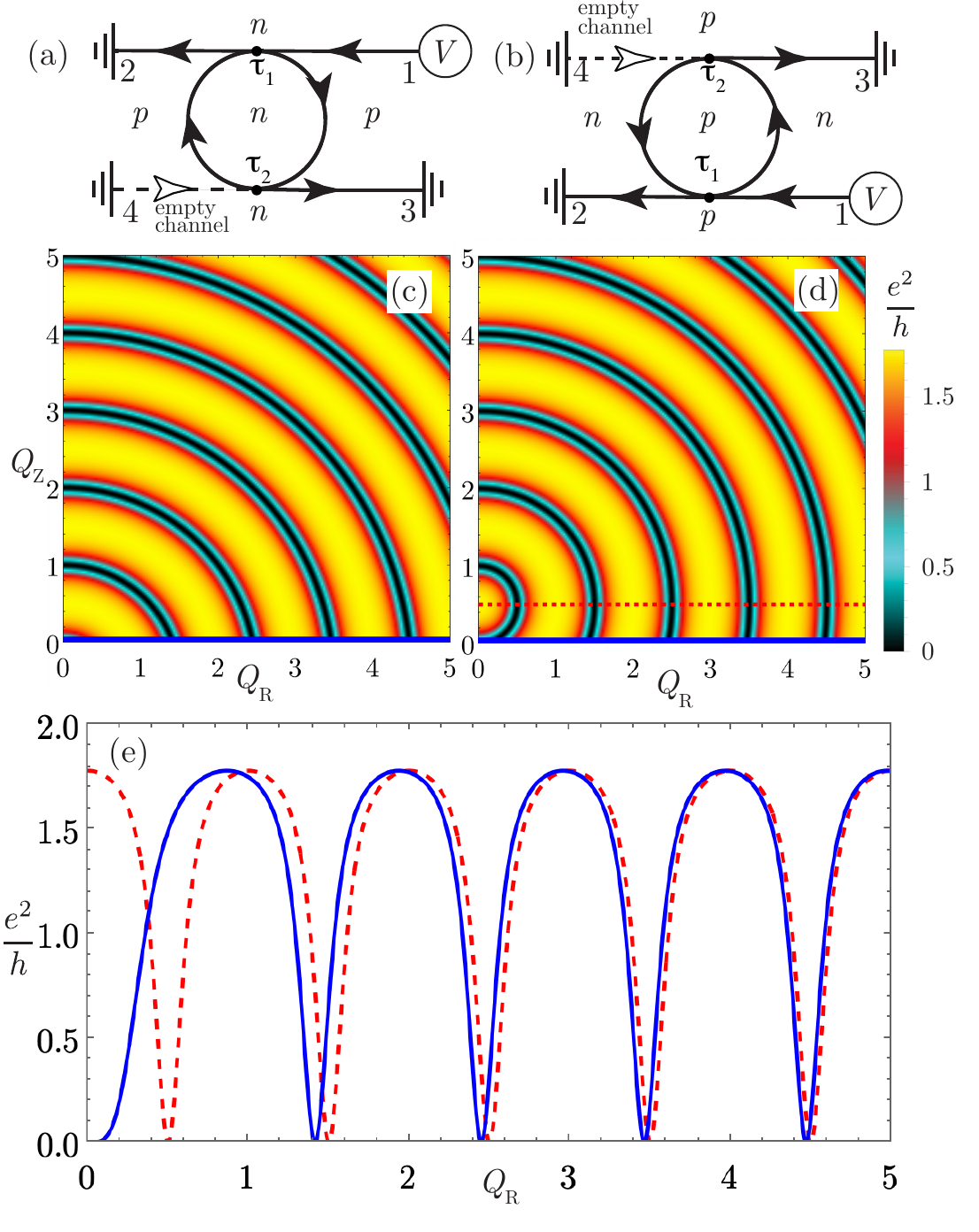}
    \caption{(a) The circuit's architecture with a $n$-doped dot. 
    (b)The same as in (a), but with opposite junctions' polarity.
    (c)-(d) Differential conductance $G_{21}$ for the circuits in (a) and (b), respectively, 
    as a function of the dimensionless Rashba and Zeeman coupling strengths. 
    (e) Cut of the differential conductance for the cases in (c) and (d) with $Q_\text{Z}=0$ (blue line) 
    and for the case in (d) with $Q_\text{Z}=1/2$ (red dashed line).}
    \label{fig3}
\end{figure}

Figure \ref{fig3} depicts the circuits' architecture built upon $n$ [Fig.~\ref{fig3}(a)] 
and $p$ [Fig.~\ref{fig3}(b)]  dots. 
Contact 1 at voltage $V$ is the carrier source, while the grounded contacts 2 and 3 act as drains. 
The grounded contact~4 contributes with an empty channel. Importantly, 
either setup can be turned into the other 
by simply inverting the $pn$~polarity, relabeling the contacts, and swapping voltages, 
meaning that a single sample could realize both interferometer in the laboratory.

Carriers injected from contact 1 propagate along a linear $pn$~junction. Traveling toward contact 2, 
they can enter the circular $pn$~junction with probability $0 < \tau_1 < 1$, 
from which they can escape at the opposite end towards contact 3 with probability $0 < \tau_2 < 1$. 
The tunnel barriers $\tau_1$ and $\tau_2$ operate as beam splitters (BSs) for the chiral modes. 
Their spin-dependent probability amplitudes are determined by projecting  
the propagating spin modes on the local basis~\cite{supplemental}.

We calculate the quantum conductance $G_{21}$ from contact 1 to contact 2 for the zero modes 
following the Landauer-B\"uttiker approach~\cite{Imry_1997,Datta1995}. 
(By unitarity, $G_{21}+G_{31}=2e^2/h$, since we are considering a single valley.) 
Obtaining the quantum transmission requires the combination of the BS scattering matrices~\cite{Datta1995},  
taking into account the spin-dependent phases $m\pi$ gathered by the carriers propagating between 
the tunnel barriers along the circular junction~\cite{supplemental}. 
These phases are obtained by setting $E_{m,s}=0$ in Eq.~\eqref{Em-ring}, 
where $m$ is not necessarily an integer for open $pn$~junctions, 
since periodic boundary conditions do not apply in the presence of contact leads. 
%
%
%
%
Figures~\ref{fig3}(c)-(e) summarize our main results. 
We plot the conductance $G_{21}$ for the two 
opposite junction polarities, as a function of dimensionless Rashba 
$Q_{\rm R}=\omega_{\rm R}/\omega_0$ and Zeeman $Q_{\rm Z}=\omega_{\rm Z}/\omega_0$ coupling strengths.   
Without loss of generality, we set $\tau_1=\tau_2=1/2$ (50\% BSs) and $\Phi \in \mathbb{N}$. 
Other settings can modify the relative amplitudes and phases of the patterns, 
but their general composition remains the same. 
We observe that the patterns in Figs.~\ref{fig3}(c) and \ref{fig3}(d) differ by a relative $\Delta Q_{\rm Z}=1$ 
shift along the Zeeman axis. 
This shift reveals significant information on the spin-state geometry of propagating channels, 
as explained below.

In Fig.~\ref{fig3}(e) we plot $G_{21}$ for $Q_{\rm Z}=0$ (solid line) and $Q_\text{Z}=1/2$ (dashed line). 
For $Q_\text{Z}=0$, the result holds for both $n$ and $p$ polarities. Here  
we find quasi-periodic oscillations as a function of $Q_{\rm R}$, 
which tend to be periodic for $Q_{\rm R} \gg 1$. 
This limit corresponds to the regime of adiabatic spin dynamics, 
where the local spin quantization axis 
is expected to point along the radial Rashba field with $\gamma \rightarrow \pi/2$ in Eq.~\eqref{eigenstates}. 
Moreover, after a round trip around the dot, the spin carriers collect a geometric phase $\varphi_{\rm g}=-\Omega/2$, 
with $\Omega= 2\pi(1-\cos \gamma)$ the solid angle subtended by the spin states on the Bloch sphere. 
In the adiabatic limit, one finds $\varphi_{\rm g} \rightarrow -\pi$.
Similar results have been reported for semiconductor Rashba rings~\cite{Frustaglia2004,Nagasawa2012}. 

The two polarities respond very differently 
to $Q_{\rm Z}$. For the $n$ dot [see Eq.~\eqref{gamma_angle} and Fig.~\ref{fig3}(c)], 
we find that $Q_{\rm Z}$ acts to the detriment of in-plane spinor polarization, 
which still requires large RSOC intensities $Q_{\rm R}$.
On the contrary, for the $p$ dot [see Eq.~\eqref{gamma_angle} and Figs.~\ref{fig3}(d) and~\ref{fig3}(e)],
at the sweet spot $Q_\text{Z}=1/2$ 
we find perfectly periodic oscillations corresponding to fully in-plane spin states ($\gamma=\pi/2$) 
regardless of the RSOC intensity, 
picking up a geometric phase $\varphi_{\rm g}=-\pi$.

\section{Discussion}\label{discussion}
All relevant features of Fig.~\ref{fig3}(d) are captured 
by a low-order semiclassical expansion of the conductance in terms of Feynman paths corresponding 
to single windings around the $p$ dot~\cite{supplemental}. In this approximation, we find 
%
%
\begin{equation}\label{G211st}
G_{21} \approx 1+ \cos\phi_\text{AB} \cos\phi_\text{S},
\end{equation}
%
%
with 
%
%
\begin{subequations}
\begin{align}
\label{ABphase}
\phi_\text{AB}&= 2\pi \Phi,\\
\label{Sphase}
\phi_\text{S}&= 2\pi \sqrt{\left(Q_{\rm Z}-\frac{1}{2}\right)^2+Q_{\rm R}^2},
\end{align}
\end{subequations}
%
%
where $\phi_\text{AB}$ and $\phi_\text{S}$ are independent phase contributions 
originating in the orbital and spin degrees of freedom, respectively. 
Equation~\eqref{G211st} reproduces well the pattern of Fig.~\ref{fig3}(d) showing circular wavefronts 
centered at $Q_\text{R}=0$, $Q_\text{Z}=1/2$. For $Q_\text{Z}=0$, 
we find from Eq.~\eqref{G211st} that 
$\phi_\text{S}=2\pi Q_\text{R} \sin \gamma - \pi \cos \gamma= 2\pi Q_\text{R} \sin \gamma -(\pi+\varphi_\text{g})$. 
This phase reduces to $\phi_\text{S} \approx 2\pi Q_\text{R}$ in the adiabatic limit $Q_\text{R} \gg 1$, 
leading to periodic oscillations of $G_{21}$ as a function of $Q_\text{R}$.
Thus, a strong RSOC drives the spin eigenstates to be in-plane, 
such that $\gamma \rightarrow \pi/2$ and $\varphi_\text{g} \rightarrow -\pi$. 
The physical realization of this formal limit is difficult in the laboratory due to the required field intensities. 
Alternatively,  we find here a shortcut by setting $Q_\text{Z}=1/2$. In this sweet spot, 
the spin phase contribution reduces exactly to $\phi_\text{S} = 2\pi Q_\text{R}$ even for weak RSOC fields, 
which assures in-plane spin eigenstates that introduce a $\pi$ phase-shift of purely geometric origin. 

We emphasize that this precise characterization of the propagating spin channels boils down to their chiral nature, in contrast to the case of semiconductor Rashba rings, 
where counter-propagating modes coexist~\cite{Frustaglia2004,Nagasawa2012,Nagasawa2013}. 
The chirality also protects the sweet spot from the effect of random impurities. 
Moreover, we expect that small deviations from a perfectly circular shape, 
breaking the rotational symmetry might induce small oscillations 
of the out-of-plane component of the spin and thus blur the sweet spot, 
but will not qualitatively alter the physics discussed here~\cite{Ying_2016}.

Finally, we briefly address the effect of the VZSOC. 
In the effective model~\eqref{effectiveHam}, it leads to a valley-dependent shift 
$\omega_\text{Z} \rightarrow \omega_\text{Z} + \tau \lambda_\text{VZ}$. Hence,
at $\omega_\text{Z}=\omega_0/2$,
the spin states~\eqref{eigenstates} will have a residual out-of-plane component, 
opposite at the two valleys. The
 valley-resolved conductances will be periodic functions of $Q_\text{R}$ only 
for $\lambda_\text{R}\gg \lambda_\text{VZ}$~\cite{supplemental}, see Eqs.~\eqref{G211st} and~\eqref{Sphase}. 
The selection of substrates inducing the weakest possible 
VZSOC~\cite{Naimer_2021,Naimer_2023} is thus essential to observing the effects described in this work.

\section{Conclusions}\label{conclusions}
We have shown that the chiral spin channels in curved graphene $pn$~junctions 
with proximitized SOCs can be precisely characterized and controlled.
We uncovered a sweet spot in the parameter space enabling an efficient manipulation 
of spin-state configurations without requiring a strong RSOC, 
which is difficult to achieve experimentally. 
This opens up new possibilities for exploring quantum-state geometry and 
advancing spintronics in graphene. 
Curved $pn$~junctions thus offer a versatile platform for investigating spin dynamics 
phenomena induced by SOCs, providing an alternative to traditional semiconductor systems.

\begin{acknowledgments} 
We thank R.~Egger and K.~Richter 
for helpful comments on the manuscript. 
D.B. acknowledges the support from the Spanish MICINN-AEI 
through Project No.~PID2020-120614GB-I00~(ENACT), the Transnational Common Laboratory $Quantum-ChemPhys$, the financial support received from the IKUR Strategy under the collaboration agreement 
between Ikerbasque Foundation and DIPC on behalf of the Department of Education of the 
Basque Government and the Gipuzkoa Provincial Council within the QUAN-000021-01 project. D.F. acknowledges support from the Spanish MICINN-AEI through 
Project No.~PID2021-127250NB-I00~(e-QSG) and from the Andalusian Government through PAIDI 2020 
Project No.~P20-00548 and FEDER Project No.~US-1380932. 
\end{acknowledgments}

\bibliography{biblio}

\begin{thebibliography}{61}
\bibitem [{sup()}]{supplemental}%
  \BibitemOpen
  \href@noop {} {}\bibinfo {note} {{S}ee {S}upplemental {M}aterial at [URL],
  which includes details of the full model and its solution in the uniform case
  and for the junction, detailed derivation of the effective 1D model,
  additional cases for the Rabi condition and the details for the construction
  of the $S$-matrix and its first order expansion.}\BibitemShut {Stop}%
\end{thebibliography}

\onecolumngrid

\pagebreak

\renewcommand{\theequation}{SE\arabic{equation}}
\renewcommand{\thefigure}{SF\arabic{figure}}
\renewcommand{\bibnumfmt}[1]{[#1]}
\renewcommand{\citenumfont}[1]{#1}
\setcounter{equation}{0}
\setcounter{figure}{0}
\setcounter{table}{0}
\setcounter{section}{0}

\begin{center}
   \large{\textbf{SUPPLEMENTAL MATERIAL}} \\ 
    \vspace{.5cm}
   \Large{\textbf{Chiral spin channels in curved graphene \emph{pn}~junctions}}
\end{center}


\section{The Hamiltonian}

In this section, we illustrate the complete low-energy Hamiltonian for a graphene monolayer
with proximitized spin-orbit couplings (SOCs). 
Following~\cite{Gmitra:2016fk} (see also~\cite{Frank_2020}), 
the full Hamiltonian reads:
\begin{equation}
    \mathcal{H} = {\mathcal H}_0 + \mathcal{H}_\Delta + \mathcal{H}_\text{Z} +
\mathcal{H}_\text{R} + \mathcal{H}_\text{KM} +  \mathcal{H}_\text{VZ}, 
\label{hamSM}
\end{equation}
where
%
%
\begin{subequations}
\begin{align}
{\mathcal H}_0 &=v_\text{F} \left( \tau_z \sigma_x \Pi_x + \sigma_y \Pi_y \right) +V, \\
\mathcal{H}_\Delta &= \Delta \sigma_z  ,\\
\mathcal{H}_\text{Z} &= \lambda_\text{Z} s_z , \\
\mathcal{H}_\text{R} &=  \frac{\lambda_\text{R}}{2} \left( \tau_z \sigma_x  s_y - \sigma_y s_x\right) 
,\\
\mathcal{H}_\text{KM} & =   \lambda_\text{KM} \tau_z \sigma_z s_z,\\ 
\mathcal{H}_\text{VZ} &= \lambda_\text{VZ}  \tau_z s_z.
\end{align}
\end{subequations}
%
%
Here, $v_\text{F}\approx10^6$\,m/s is the graphene's Fermi velocity, 
${\bm \Pi}= -i \hbar \nabla + \frac{e}{c}{\bf A}$ the kinetic momentum,
$\mathbf{A}=\frac{B}{2}(-y,x,0)$ the vector potential in the symmetric gauge 
(we assume $B>0$), and $V({\bf r})$ the potential defining the circular $pn$~junction. 
The symbols $\bm \tau$/$\bm \sigma$/$\bf s$  denote the
valley/sublattice/spin Pauli matrices.
The Hamiltonian~\eqref{hamSM} is diagonal in valley space. It includes the 
sublattice-symmetry breaking term $\mathcal{H}_\Delta$,
the Zeeman term  $\mathcal{H}_\text{Z}$ 
(with  $\lambda_\text{Z}= \frac{g_s  \mu_B}{2}B$),
the Kane and Mele (or intrinsic) SOC $\mathcal{H}_\text{KM}$, the Rashba SOC $\mathcal{H}_\text{R}$,
and the valley-Zeeman SOC  $\mathcal{H}_\text{VZ}$. 
For completeness, we include the sublattice-symmetry breaking term 
$\mathcal{H}_\Delta$, which we have neglected in the main text.

The wave function is an $8$-component spinor
\begin{equation}
 \left( \Psi_{\text{A}\uparrow},\Psi_{\text{B}\uparrow},
\Psi_{\text{A}\downarrow},\Psi_{\text{B}\downarrow},\Psi'_{\text{A}\uparrow},\Psi'_{\text{B}\uparrow},
\Psi'_{\text{A}\downarrow},\Psi'_{\text{B}\downarrow}\right),
\end{equation}
where the unprimed and primed components are the amplitudes at the valley $K$ ($\tau_z=+1$) 
and $K'$ ($\tau_z=-1$), respectively.
The Hamiltonian $\mathcal{H}$ is invariant under the time-reversal operation 
$\mathcal{T}=is_y\tau_x \mathcal{K}$ up to the inversion of the magnetic field:
\begin{equation}
  \mathcal{T} \, \mathcal{H}(\mathbf{B}) \, \mathcal{T}^\dagger = \mathcal{H}(-\mathbf{B}),
\end{equation}
and commutes with the total angular momentum operator $J= L_z +\frac{1}{2}\left(\tau_z \sigma_z+s_z\right)$.

Since $\mathcal{H}$ is diagonal in valley space, we will focus on a 
single valley ($\tau_z=+1$) and omit the valley index. 
Then, the wave function $\Psi$ is a four-component spinor in sublattice/spin space,
$\Psi^\text{T} = \left( \Psi_{\text{A}\uparrow},\Psi_{\text{B}\uparrow},
\Psi_{\text{A}\downarrow},\Psi_{\text{B}\downarrow}\right)$.
The single-valley Hamiltonians are related by the unitary transformation
%
%
\begin{equation}
\mathcal{H}_{\tau_z=-1}(\Delta, \lambda_\text{VZ}) = 
 i\sigma_y {\cal H}_{\tau_z=+1}(-\Delta,-\lambda_\text{VZ}) (-i\sigma_y).
\end{equation}
%
%
Using this identity, one can find the eigenstates at the valley $\tau_z=-1$ once 
the eigenstates at the valley $\tau_z=+1$ are determined.

Before closing this section, we notice that we express energy in units of the relativistic 
cyclotron energy $\hbar \omega_c$, length in units of the magnetic length $\ell_B$, 
and wave vectors in units of $\ell_B^{-1}$, with
%
%
\begin{align*}
& \ell_B =\sqrt{\frac{\hbar c}{eB}} = \frac{26}{\sqrt{B[\text{T}]}}\, \text{nm},  \quad  
\hbar \omega_c  =\frac{\hbar v_\text{F}}{\ell_B} = 26\sqrt{B[\text{T}]}\, \text{meV}, \quad 
\hbar \omega_\text{Z} = \frac{g_s\mu_B}{2} B = 5.8\times 10^{-2} B[\text{T}] \, \text{meV}.
\end{align*}
%
%
We set $e=\hbar=v_\text{F}=1$ unless specified otherwise.

\section{Exact model solution}
\label{Gensol}

In this section, we provide the exact solution of the problem of graphene's Landau levels 
in the symmetric gauge in the presence of SOCs and a constant potential.
(See \cite{De_Martino_2011,Bercioux_2019} for the solution to this problem in the Landau gauge.)
Since we work in a given valley, the valley-Zeeman term can be absorbed into 
the Zeeman term and will be omitted below. The single-valley Hamiltonian ($\tau_z=+1$) 
in the symmetric gauge commutes with the total angular momentum 
\begin{equation}
    J=L_z+\frac{\sigma_z}{2}+\frac{s_z}{2},
\end{equation}
with $L_z = -i\partial_\theta$, hence the eigenfunctions can be 
labeled by the eigenvalues of $J$, which span the set of integers, 
and take the form 
\begin{equation}
    \Psi_j(\mathbf{r}) = \frac{e^{i(j-\frac{\sigma_z}{2}-\frac{s_z}{2})\theta}}{\sqrt{2\pi}} \psi_j(r),
\end{equation}
where $(r,\theta)$ are polar coordinates and $j\in \mathbb{Z}$. 
The radial spinor $\psi_j(r)$ is a solution of the equation
%
%
\begin{equation}
    \left( \mathcal{H}_j-E \right) \psi_j=0,
    \label{Diraceq}
\end{equation}
%
%
where 
%
%
\begin{equation}
    \mathcal{H}_j = 
    e^{-i(j-\frac{\sigma_z}{2}-\frac{s_z}{2})\theta} \, \mathcal{H} 
    \,e^{i(j-\frac{\sigma_z}{2}-\frac{s_z}{2})\theta}
\end{equation}
%
%
is the radial Hamiltonian in a fixed $j$ sector:
\begin{align}
{\cal H}_j-E= \begin{pmatrix}
-\mu_+ & -i( \frac{d}{dr} + \frac{j}{r} +  \frac{r}{2}) & 0 & 0 \\
 i( -\frac{d}{dr} + \frac{j-1}{r} + \frac{r}{2})   & -\mu_- & -i\lambda_\text{R} & 0 \\
0 &i \lambda_\text{R} & -\nu_-  & -i( \frac{d}{dr} + \frac{j+1}{r} + \frac{r}{2} ) \\
0 & 0 & i( -\frac{d}{dr} + \frac{j}{r} +\frac{r}{2} ) & - \nu_+ 
\end{pmatrix}.
\label{hamjrad}
\end{align}
Here, we have introduced the auxiliary symbols
%
%
\begin{subequations}\label{munu}
\begin{align}
\mu_\pm& = E-V-\left( \lambda_\text{Z}  \pm \lambda_\text{KM} \pm \Delta \right),\label{mupm}  
\\
\nu_\pm& =  E-V-\left( - \lambda_\text{Z}   \pm \lambda_\text{KM} \mp \Delta \right), \label{nupm}
\end{align}
\end{subequations}
%
%
and we will use the notation
%
%
\begin{align}
  \mu&=\mu_+\mu_- =(E - V - \lambda_\text{Z} )^2- (\lambda_{\text{KM}} + \Delta)^2,\\ 
  \nu&=\nu_+\nu_-=(E - V + \lambda_\text{Z} )^2- (\lambda_{\text{KM}} - \Delta)^2.  
\end{align}
%
%
In terms of the variable $\xi=r^2/2$, we find
%
%
\begin{align}\label{ham_bulk}
\mathcal{H}_j-E= 
\begin{pmatrix}
-\mu_+ &- i\sqrt{2\xi} \left( \frac{d}{d\xi} + \frac{1}{2} + \frac{j}{2\xi}  \right) & 0 & 0 \\
i\sqrt{2\xi} \left( -\frac{d}{d\xi} + \frac{1}{2} + \frac{j-1}{2\xi} \right)   &
- \mu_- & - i\lambda_\text{R} & 0 \\
0 & i\lambda_\text{R} & -\nu_-  & - i\sqrt{2\xi} \left( \frac{d}{d\xi} + 
\frac{1}{2}+ \frac{j+1}{2\xi} \right) \\
0 & 0 & i\sqrt{2\xi} \left(- \frac{d}{d\xi} + \frac{1}{2} + \frac{j}{2\xi} \right) &- \nu_+ 
\end{pmatrix}.
\end{align}
%
%
The general solution of Eq.~\eqref{Diraceq} can be expressed in terms of confluent 
hypergeometric functions~\cite{Olver_2010}. In the following, we will present the solutions 
separately for $j \geq 0$ and $j<0$.

\subsection{Case $j \geq 0$}
\label{j>0}

First, we assume $j>0$. The solutions of graphene's Landau levels problem without SOCs 
(see, e.g., \cite{DeMartino_2007,DeMartino_2010,Cohnitz_2016}) 
and with SOCs in the Landau gauge \cite{De_Martino_2011,Bercioux_2019}
suggest the following ansatz:
%
%
\begin{align}
\psi_j (\xi) =e^{-\xi/2} \xi^{j/2}
\begin{pmatrix}
d_1 \xi^{-1/2}  M(a,j,\xi) \\
i  d_2 M(a,j+1,\xi) \\
  d_3 M(a,j+1,\xi) \\
id_4 \xi^{1/2}  M(a,j+2,\xi)
\end{pmatrix},
\label{Mjpos}
\end{align}
%
%
where $d_i$ are constant coefficients (for simplicity, we omit the index $j$ on the coefficients), 
and $M(a,b,\xi)$ denotes the confluent hypergeometric function of the first kind~\cite{Olver_2010}, 
regular at the origin. The parameter $a$ will be determined below.
By using recurrence relations between confluent hypergeometric functions, 
Eq.~\eqref{ham_bulk} is converted into a linear system for the coefficients $d_i$:
%
%
\begin{align}\label{coeff_d_i_y_m_0}
\begin{pmatrix}
- \mu_+ & \sqrt{2}j & 0 & 0 \\
\sqrt{2}\left(1-\frac{a}{j}\right)  & - \mu_- & -\lambda_\text{R} & 0 \\
0 & -\lambda_\text{R} & - \nu_-  & \sqrt{2}(j+1) \\
0 & 0 & \sqrt{2} \left(1-\frac{a}{j+1} \right) & - \nu_+ 
\end{pmatrix}
\begin{pmatrix}
d_1 \\ d_2 \\ d_3 \\ d_4
\end{pmatrix}= 0.
\end{align}
%
%
The existence of a non-trivial solution requires 
the vanishing of the determinant of the coefficient matrix:
%
%
\begin{equation}\label{se_j_gz}
 [2(a-j)+\mu][2(a-j-1)+\nu] -\lambda_\text{R}^2 \mu_+\nu_+=0.
\end{equation}
%
%
The solution of the linear system~\eqref{coeff_d_i_y_m_0} is (up to an overall constant) 
%
%
\begin{align}
\begin{pmatrix}
d_1 \\ d_2 \\ d_3 \\ d_4
\end{pmatrix}
=\begin{pmatrix}
\sqrt{2}j \\  
\mu_+ \\  
- \frac{2(a-j)+\mu}{\lambda_\text{R}} \\
 \frac{\sqrt{2}(a-j-1)[2(a-j)+\mu])}{\lambda_\text{R}\nu_+(j+1)}
\end{pmatrix}.
\label{dMjpos}
\end{align}
%
%

A second solution, singular at the origin, is built using the confluent 
hypergeometric function of the second kind $U(a,b,\xi)$~\cite{Olver_2010}:
%
%
\begin{align}
\psi_j (\xi) =e^{-\xi/2} \xi^{j/2}
\begin{pmatrix}
d_1 \xi^{-1/2} U(a,j,\xi) \\ 
i d_2 U(a,j+1,\xi) \\  
d_3 U(a,j+1,\xi) \\ 
i  \xi^{1/2} d_4 U(a,j+2,\xi)
\end{pmatrix}.
\label{Ujpos}
 \end{align}
%
%
The corresponding linear system for the coefficients $d_i$ is
%
%
\begin{align}
\begin{pmatrix}
-\mu_+ & \sqrt{2}(j-a) & 0 & 0 \\
\sqrt{2} & - \mu_- & -\lambda_\text{R} & 0 \\
0 & - \lambda_\text{R} & - \nu_-  & \sqrt{2}(j+1-a) \\
0 & 0 & \sqrt{2} & - \nu_+ 
\end{pmatrix}
\begin{pmatrix}
d_1 \\ d_2 \\ d_3 \\ d_4
\end{pmatrix} = 0,
\end{align}
%
%
and the determinant equation is the same as in Eq.~\eqref{se_j_gz}.  
The solution of this linear system gives
%
%
\begin{align}
\begin{pmatrix}
d_1 \\ d_2 \\ d_3 \\ d_4
\end{pmatrix}
=\begin{pmatrix}
\sqrt{2}(j-a) \\
\mu_+ \\
-\frac{2(a-j)+\mu}{\lambda_\text{R}} \\
-\frac{\sqrt{2}[2(a-j)+\mu]}{\lambda_\text{R} \nu_+}
\end{pmatrix}.
\label{dUjpos}
\end{align}
%
%

The solution for $j=0$ can now be obtained by taking the 
limit $j\rightarrow 0$ in the previous formulas and using 
the following identities:
\begin{align}
\lim_{j\to 0} j M(a,j,\xi) &= a \xi M(a+1,2,\xi) ,\\
U(a,0,\xi) &= \xi U(a+1,2,\xi).
\end{align}

\subsection{Case $j < 0$}
\label{j<0}

In this case, the correct ansatz for the solution regular at the origin reads
%
%
\begin{align}
\psi_j(\xi) =e^{-\xi/2} \xi^{-j/2}
\begin{pmatrix}
d_1 \xi^{1/2} M(a+1,-j+2  ,\xi) \\
i  d_2M(a,-j+1 ,\xi) \\
  d_3M(a,-j+1,\xi)\\
i d_4 \xi^{-1/2} M(a-1, -j,\xi))
\end{pmatrix}.
\label{Mjneg}
\end{align}
%
%
Then, the algebraic equation for the coefficients $d_i$ is
%
%
\begin{align}
\label{spinor_jm0}
\begin{pmatrix}
-\mu_+ & \sqrt{2} \frac{a}{1-j} & 0 & 0 \\
\sqrt{2}(j-1) & -\mu_- & -\lambda_\text{R} & 0 \\
0 & - \lambda_\text{R} & - \nu_-  & \sqrt{2}\frac{1-a}{j} \\
0 & 0 & \sqrt{2}j & - \nu_+ 
\end{pmatrix}  
\begin{pmatrix}
d_1 \\ d_2 \\ d_3 \\ d_4
\end{pmatrix} = 0.
\end{align}
%
%
The condition of vanishing determinant reads
%
%
\begin{equation}\label{se_j_sz}
(2a+ \mu)(2a - 2 +\nu)-\lambda_\text{R}^2 \mu_+\nu_+ =0,
\end{equation}
%
%
and the solution of the linear system (up to an overall constant) is 
%
%
\begin{align}
\begin{pmatrix}
d_1 \\
d_2 \\
d_3 \\
d_4
\end{pmatrix} =
\begin{pmatrix} 
\frac{\sqrt{2}\lambda_\text{R} \nu_+ a}{1-j} \\
\lambda_\text{R} \mu_+\nu_+ \\ 
 -\nu_+(2a+\mu)  \\
 -\sqrt{2}j(2a+\mu)
\end{pmatrix}.
\label{dMjneg}
\end{align}
%
%

The second solution, singular at the origin, is given by
%
%
\begin{align}
\psi_j(\xi) =e^{-\xi/2} \xi^{-j/2}
\begin{pmatrix}
d_1 \xi^{1/2} U(a+1,-j+2,\xi ) \\
i d_2  U(a,-j+1,\xi) \\
d_3  U(a,-j+1,\xi) \\
i d_4 \xi^{-1/2} U(a-1, - j ,\xi)
\end{pmatrix}.
\label{Ujneg}
\end{align}
%
%
The associated linear system is
%
%
\begin{align}
\begin{pmatrix}
-\mu_+ & -\sqrt{2}a& 0 & 0 \\
\sqrt{2} & -\mu_- & - \lambda_\text{R} & 0 \\
0 & -\lambda_\text{R} & -\nu_-  & \sqrt{2}(1-a) \\
0 & 0 & \sqrt{2} & -\nu_+ 
\end{pmatrix}
\begin{pmatrix}
d_1 \\ d_2 \\ d_3 \\ d_4
\end{pmatrix} =0,
\end{align}
%
%
with the same determinant equation as in Eq.~\eqref{se_j_sz}, and the solution given by
%
%
\begin{align}
\begin{pmatrix}
d_1 \\ d_2 \\ d_3 \\ d_4
\end{pmatrix}=
\begin{pmatrix}
-\sqrt{2} \lambda_\text{R} \nu_+ a  \\ 
\lambda_\text{R}\mu_+ \nu_+ \\
- \nu_+ (2a+\mu) \\
- \sqrt{2}(2a+\mu)
\end{pmatrix}.
\label{dUjneg}
\end{align}
%
%

We note in passing that, by taking the limit $j\rightarrow 0$ in the
formulas above, we recover the solution for $j=0$ given in Sec.~\ref{j>0}.

\subsection{General solution}

The two determinant equations \eqref{se_j_gz} and \eqref{se_j_sz} 
can be merged into a single equation:
%
%
\begin{equation}\label{deteq}
 [2(a-j\Theta(j))+\mu][2(a-j\Theta(j)-1)+\nu] -\lambda_\text{R}^2 \mu_+\nu_+=0,
\end{equation}
%
%
where $\Theta(x)$ is the Heaviside function. (We adopt the convention $\Theta(0)=1$.)
This condition admits the solutions $a=a_\pm$ given by
%
%
\begin{equation}\label{det_eq_one}
a_\pm = j\Theta(j) - \frac{1}{4} \left[
\mu+\nu-2 \pm \sqrt{(\mu-\nu+2)^2+4\lambda_\text{R}^2 \mu_+\nu_+} \right].
\end{equation}
%
%
We denote by $\psi^<_j(\xi)$ the wave functions regular at the origin, 
Eqs.~\eqref{Mjpos} and \eqref{Mjneg},
and by $\psi^>_j(\xi)$ the wave functions singular at the origin, 
Eqs.~\eqref{Ujpos} and \eqref{Ujneg}.
The eigenspace of energy $E$ and total angular momentum~$j$ is then spanned by 
the linear combinations of the four solutions obtained by taking $\psi^<_j$ and
$\psi^>_j$ with $a=a_\pm$ in Eq.~\eqref{det_eq_one}:
$$
\psi_j(\xi) = c_1 \psi^<_{j,a_+}(\xi) +  c_2 \psi^<_{j,a_-}(\xi) + 
c_3 \psi^>_{j,a_+}(\xi) + c_4 \psi^>_{j,a_-}(\xi).
$$
As we will see below, the quantized energy eigenvalues (Landau levels) 
are obtained by imposing appropriate conditions on this general solution. 
In Sec.~\ref{LL} we consider the case of a uniform system, where
the quantization condition originates simply from the requirement of 
normalizability. In Sec.~\ref{LLpn} we consider the case of a 
circular $pn$~junction, where the quantization condition arises 
from the combination of the requirements of normalizability and 
continuity of the wave function.

\section{Uniform system}
\label{LL}

For a uniform system ($V=0$) we have to select the solutions regular 
at the origin, Eqs.~\eqref{Mjpos} and \eqref{Mjneg}, 
which involve the functions $M(a,b,\xi)$. Then normalizability requires 
that the first argument of $M$ is a non-positive integer $-n$, where $n$ is 
interpreted as the radial quantum number. As a result, we find 
two sets of Landau levels, obtained by solving the equations
\begin{equation}
\begin{cases}
a_\pm(E) = - n & n=0,1,2, \dots, \; \text{for} \;  j \geq 0 \\
a_\pm(E)-1 = - n & n=0,1,2, \dots, \; \text{for} \;   j < 0
\end{cases},
\end{equation}
with $a_\pm$ in Eq.~\eqref{det_eq_one}.
For the correct counting of the solutions, one should notice the following:
\begin{itemize}

\item Case $j= 0$: for $n=0$ (i.e., $a=0$) the solution of Eq.~\eqref{deteq} 
with $\mu_+=0$ must be omitted, because all wave function components vanish, 
see Eq.~\eqref{dMjpos}.

\item Case $j<0$: for $n=0$ (i.e., $a=1$) only the solution of Eq.~\eqref{deteq} with 
$\nu_+=0$ is allowed because the first three components of the wave function 
(which are not normalizable functions if $a=1$) 
have a vanishing coefficient, see Eq.~\eqref{dMjneg}. 
For $n=1$ (i.e., $a=0$) the solution of Eq.~\eqref{deteq} with $\mu_+=0$ 
must be omitted for the same reason as in the case $j=0$ above.

\end{itemize}

The quantization equation~\eqref{deteq} is quartic in the energy and can be solved explicitly. 
However, the general expression of the solutions is cumbersome and not particularly illuminating. 
Below, we briefly discuss few special cases and give the explicit formulas for  
the corresponding energy eigenvalues.

\begin{itemize}

\item 
If the Rashba SOC vanishes, Eq.~\eqref{deteq} decouples into two separate equations, 
each giving a set of spin-polarized Landau levels. From $2(a-j\Theta(j))+\mu =0$ we find
the spin-up levels 
%
%
\begin{align}\label{LL_zeroRashbaup}
E_{n,j,\alpha,-}=\lambda_\text{Z}   +\alpha \sqrt{2(n+j\Theta(j))+
(\lambda_\text{KM}+\Delta)^2}, \quad  n=0,1,2, \dots, \quad 
j=0, \pm 1,\pm 2, \dots, \quad \alpha=\pm 1, 
\end{align}
%
%
where for $n=0$ and $j\leq 0$ only the level $E_{0,j}=\lambda_\text{Z}-\lambda_\text{KM}-\Delta$ 
must be kept. 
From $2(a-j\Theta(j)-1)+\nu=0$ we find the spin-down levels 
%
%
\begin{align}\label{LL_zeroRashbadown}
E_{n,j,\alpha,+}= - \lambda_\text{Z} +
\alpha \sqrt{2(n+(j+1)\Theta(j))+(\lambda_\text{KM}-\Delta)^2}, 
\quad n=0,1,2, \dots, \quad 
j=0, \pm 1,\pm 2, \dots,\quad \alpha = \pm 1,
\end{align}
%
%
where for $n=0$ and $j< 0$ only the level $E_{0,j}= - \lambda_\text{Z}+\lambda_\text{KM}-\Delta$ 
must be kept. 
For $\lambda_\text{Z}=\lambda_\text{KM}=\Delta=0$, the expressions in Eqs.~\eqref{LL_zeroRashbaup} 
and~\eqref{LL_zeroRashbadown} coincide with the Landau level formula in the symmetric gauge 
(see, e.g.,~\cite{Cohnitz_2016}) after the replacement $j =j' \pm \frac{1}{2}$ 
for spin up/down states, with $j'$ half-integer. 

%
%
\begin{figure}
    \centering
    \includegraphics[width=\textwidth]{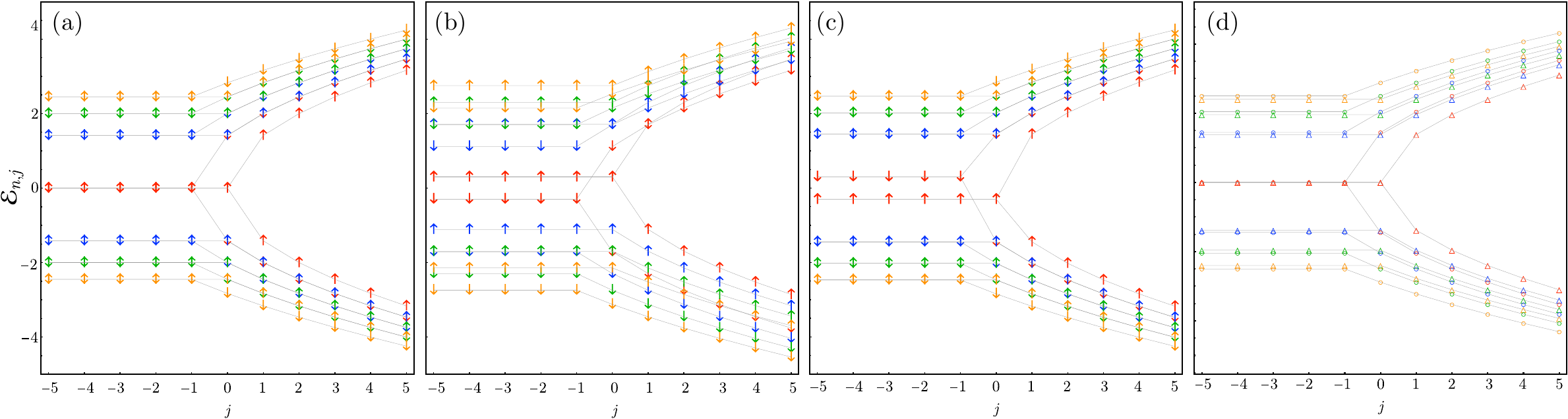}
    \caption{Landau level spectrum in the symmetric gauge. 
    (a) Spin degenerate case, with $\lambda_\text{Z}=\lambda_\text{KM}=\lambda_\text{R}=\Delta=0$.
    (b) Case with only Zeeman splitting, with $\lambda_\text{Z}=0.3$. 
    (c) Case with only intrinsic SOC, with $\lambda_\text{KM}=0.3$. 
    (d) Case with only Rashba SOC, with $\lambda_\text{R}=0.3$. 
    In panels (a) to (c), the arrows are associated with the eigenvalues of $s_z$. 
    In all panels, the color refers to 
    the radial quantum number $n$: red to $n=0$, blue to $n=1$, 
    green to $n=2$ and orange to $n=3$.}
    \label{SM_fig_1}
\end{figure}
%
%

\item
If we have only a finite Rashba SOC and all the other couplings are set to zero, 
the Landau levels obtained by solving Eq.~\eqref{deteq} are given by (see also \cite{Bercioux_2019})
%
%
\begin{equation}
E_{n,j,\alpha,\beta}= \alpha \left[ 2[n+(j+1)\Theta(j)] -1 + \frac{\lambda_\text{R}^2}{2}
+\beta \sqrt{\left(1-\frac{\lambda^2_\text{R}}{2}\right)^2+2 \lambda_\text{R}^2 [n+(j+1)\Theta(j)] }
\right]^{\frac{1}{2}}, \quad \alpha,\beta=\pm 1.
\end{equation}
%
%
For the correct counting of the states, one should keep in mind 
the remarks at the beginning of this section.
It is interesting to observe that at the lowest order in $\lambda_\text{R}$
one finds 
%
%
\begin{align}
\label{perturbativeR}
E_{n,j,\alpha,\beta} \approx 
\alpha \left( 1 + \beta \frac{\lambda_\text{R}^2}{2}\right) \sqrt{2[n+(j+1)\Theta(j)]+\beta-1} .
\end{align}
%
%
We see that the effect of a small Rashba SOC is essentially a renormalization 
of the cyclotron frequency. We note in passing that the states corresponding to the two sets of levels in 
Eq.~\eqref{perturbativeR} are not eigenstates of $s_z$.
\end{itemize}

We illustrate in Fig.~\ref{SM_fig_1} the exact Landau level spectrum 
in the symmetric gauge for four relevant cases. 
We observe that, in all cases, at a fixed value of the radial quantum number $n$,  
the energy is independent of the angular quantum number $j$ for $j\leq n$ or $j\leq n-1$. 
In the first three panels, we set $\lambda_\text{R}=0$, hence 
the spin projection in the $z$ direction is a good quantum number
and the eigenfunctions describe spin states polarized along the $z$ axis.
In panel~\ref{SM_fig_1}(a) we present the spin degenerate case with 
$\lambda_\text{Z}=\lambda_\text{KM}=\lambda_\text{R}=\Delta=0$. The spectra 
of spin-up and spin-down states appear to have a relative horizontal shift, 
because we label  our states with the total angular momentum $j$. They coincide if we plot
the spin-up spectrum versus $j'=j-\frac{1}{2}$ and the spin-down spectrum versus $j'=j+\frac{1}{2}$.
This shift is the reason why the lowest-energy states with $j\geq0$ appear singly degenerate. 
If only the Zeeman coupling is active, see panel~\ref{SM_fig_1}(b),
we observe the usual energy shift, 
upwards for spin-up states and downwards for spin-down states.
In the case that only the intrinsic SOC is active, illustrated in panel~\ref{SM_fig_1}(c), 
we see that the spin degeneracy of the zero-energy Landau level is lifted, 
while all other levels remain spin degenerate. 

Finally, in panel~\ref{SM_fig_1}(d) we show the spectrum when only 
the Rashba SOC is active. In this case, the projection of the spin along the $z$ axis 
is no longer a good quantum number, because the SOC mixes spin-up and spin-down 
states. As a result, the spin degeneracy of all levels is lifted, 
with the exception of the zero-energy level, which remains doubly degenerate at zero energy.
This residual degeneracy is a result of the fact that the zero-energy states have support 
on a single sublattice~\cite{De_Martino_2011,Bercioux_2019}. 

%
%
\begin{figure}[!t]
    \centering
    \includegraphics[width=0.8\textwidth]{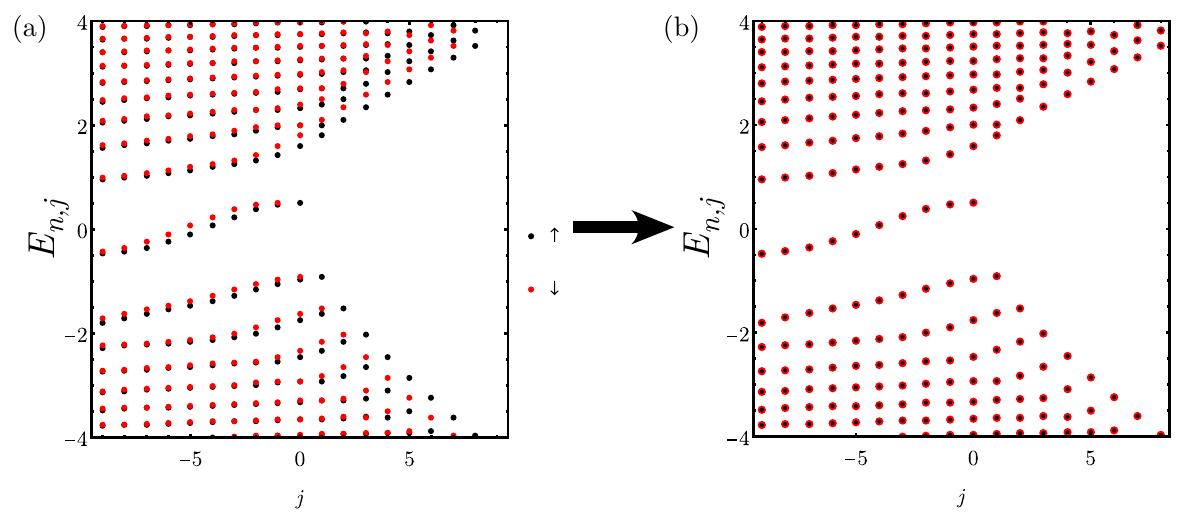}
    \caption{(a) Spectrum of the circular $pn$~junction with $\xi_0=5.1$ and $V_0=0.51$
    as a function of $j$ for vanishing SOCs and Zeeman coupling.
    (b) The same as in (a) but with the spin-down states translated by $1$  
    along the $j$-axis, showing that they coincide, as expected because of spin degeneracy.}
    \label{SM_Fig_2}
\end{figure}
%
%
\section{Landau levels in a \emph{pn}~junction}
\label{LLpn}

Next, we discuss the exact solution of the Landau level problem in the case 
of a $pn$~junction. We assume that the potential has the following profile:
%
%
\begin{align}\label{pn_potential}
V(r) = V_0\, \text{sign}(R-r), \quad V_0>0,
\end{align}
%
namely,  $V=V_0$ within a disc of radius $R$ and $V=-V_0$ outside the disc. We use $\xi_0=R^2/2$ 
as a measure for the size of the circular junction. Using
the solutions found in Sec.~\ref{Gensol}, we write the radial wave function as
%
%
\begin{align}\label{wf_interface}
\psi(\xi) =  \begin{cases}
 c_1 \psi^<_{a_+,-V_0} (\xi) +c_2 \psi^<_{a_-,-V_0} (\xi)  & \xi<\xi_0 \\ 
 c_3\psi^>_{a_+,V_0} (\xi) +c_4 \psi^>_{a_-,V_0} (\xi)  & \xi>\xi_0  
\end{cases} .
\end{align}
%
%
Here we omit the index $j$, being understood that we work at 
fixed angular momentum, and we append two indexes to indicate the values of the parameter 
$a$ and of the potential $V$. The eigenenergies and the eigenstates are obtained 
by matching the wave functions at the boundary of the disc $\xi=\xi_0$:
%
%
\begin{align}
c_1 \psi^<_{a_+,-V_0} (\xi_0) +c_2 \psi^<_{a_-,-V_0} (\xi_0) = c_3 \psi^>_{a_+,V_0} (\xi_0) + c_4 \psi^>_{a_-,V_0} (\xi_0).
\label{linearsym}
\end{align}
%
%
In analogy to the case of a linear junction~\cite{Cohnitz_2016,Bercioux_2019}, 
we obtain a linear system for the $c_i$, with the matrix of coefficients given by
%
%
\begin{align}
{\bf W} = \begin{bmatrix} \psi^<_{a_+,-V_0} (\xi_0) &  \psi^<_{a_-,-V_0} (\xi_0)  
& - \psi^>_{a_+,V_0} (\xi_0)  & -\psi^>_{a_-,V_0} (\xi_0) \end{bmatrix}.
\end{align}
%
%
The allowed energy eigenvalues are found by solving the equation 
%
%
\begin{align}\label{spectral_equation} 
\det \mathbf{W}  = 0.
\end{align}
Once the eigenvalues are determined, the corresponding normalized eigenstates 
can be calculated from Eq.~\eqref{wf_interface} using the solution of the
linear system~\eqref{linearsym}.

In Fig.~\ref{SM_Fig_2} we show the exact spectrum of the circular $pn$~junction 
obtained from the numerical solution of Eq.~\eqref{spectral_equation} in the 
absence of Rashba SOC for the spin-degenerate case.  The effect of the potential step is that
the levels acquire a dispersion in $j$. As observed in the 
uniform case discussed in Sec.~\ref{LL}, the spectra for spin-up and spin-down states 
appear horizontally shifted one with respect to the other, 
which results from labeling the states with the total angular momentum $j$. 
As shown in panel~\ref{SM_Fig_2}(b), when the spin-down spectrum is shifted 
to the right by $1$, they do overlap. This observation suggests defining
the splitting of the energy levels as the difference 
$|E_{n,j,\alpha,+}-E_{n,j-1,\alpha,-}|$, which vanishes in the spin-degenerate case.
%
%
%
\begin{figure*}
    \centering
    \includegraphics[width=0.9\textwidth]{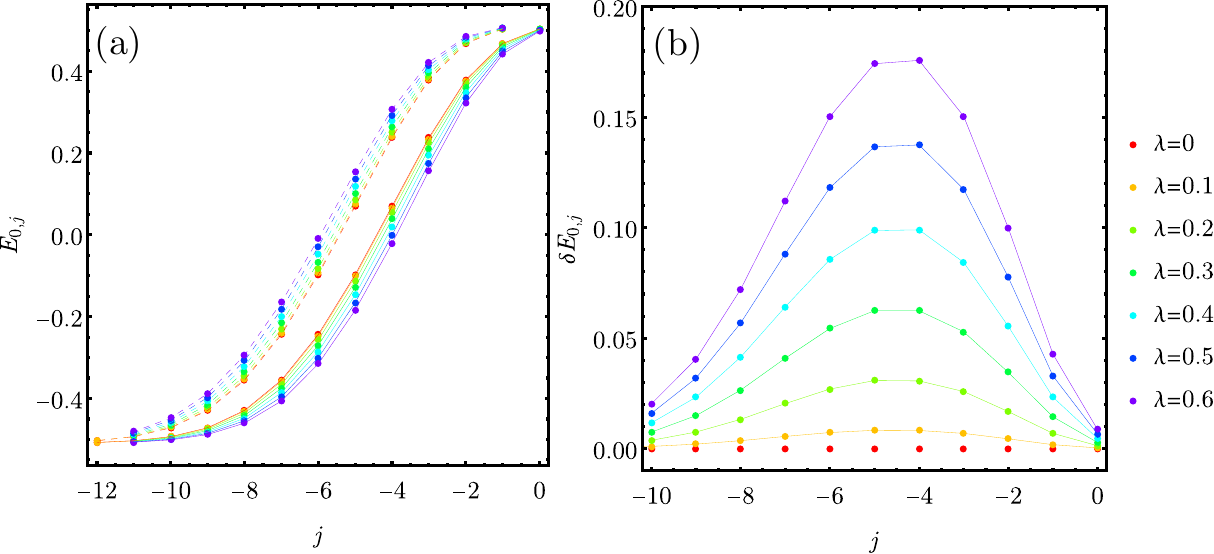}
    \caption{(a) The energy spectrum of the zero modes as a function of the total angular momentum $j$ for several values of the Rashba SOC. (b) The zero-mode 
    splitting defined in Eq.~\eqref{spinsplitting} as a function of the total angular momentum $j$ for several values of the Rashba SOC. For both panels, the values of the Rashba SOC are indicated on the side of the panel (b).}
    \label{SM_Fig_3}
\end{figure*}
%
%

We now focus on the two lowest-energy levels, which we refer to as the top and bottom modes and  
denote as $E^\text{T}_{0,j}$ and $E^\text{B}_{0,j}$ with corresponding radial wave functions
$\psi^\text{T}_j(\xi)$ and $\psi^\text{B}_j(\xi)$. In Fig.~\ref{SM_Fig_3}(a) we show their 
dispersion as a function of the total angular momentum $j$ for different values of the Rashba SOC.
We see that the $j$-dependence around zero energy is linear with good approximation.
In Fig.~\ref{SM_Fig_3}(b) we show the energy splitting of the zero modes, defined as
%
%
\begin{equation}\label{spinsplitting}
    \delta E_{0,j}=|E^\text{T}_{0,j}-E^\text{B}_{0,j-1}|.
\end{equation}
%
%

We observe that $\delta E_{0,j}$ is not constant but depends quite strongly 
on the value of the angular momentum $j$.
This dependence can be rationalized by considering the Rashba SOC
as a perturbation. As shown below, the spatial location of the zero modes 
is essentially determined by $j$. For $j=0$ and $j\gg 1$, the radial wave function 
is localized far from the $pn$~interface. In this case, the zero modes are supported on 
only one of the sublattices, so the Rashba SOC matrix element is very small. 
For values of $j$ at which the wave function is localized close to the $pn$~interface, 
instead, the zero modes have support on both sublattices so that
the matrix element of the Rashba SOC is largest and the splitting reaches a maximum.
A similar effect was observed in the case of the linear $pn$~junction~\cite{Bercioux_2019}.
%
%
\begin{figure*}
    \centering
    \includegraphics[width=0.85\textwidth]{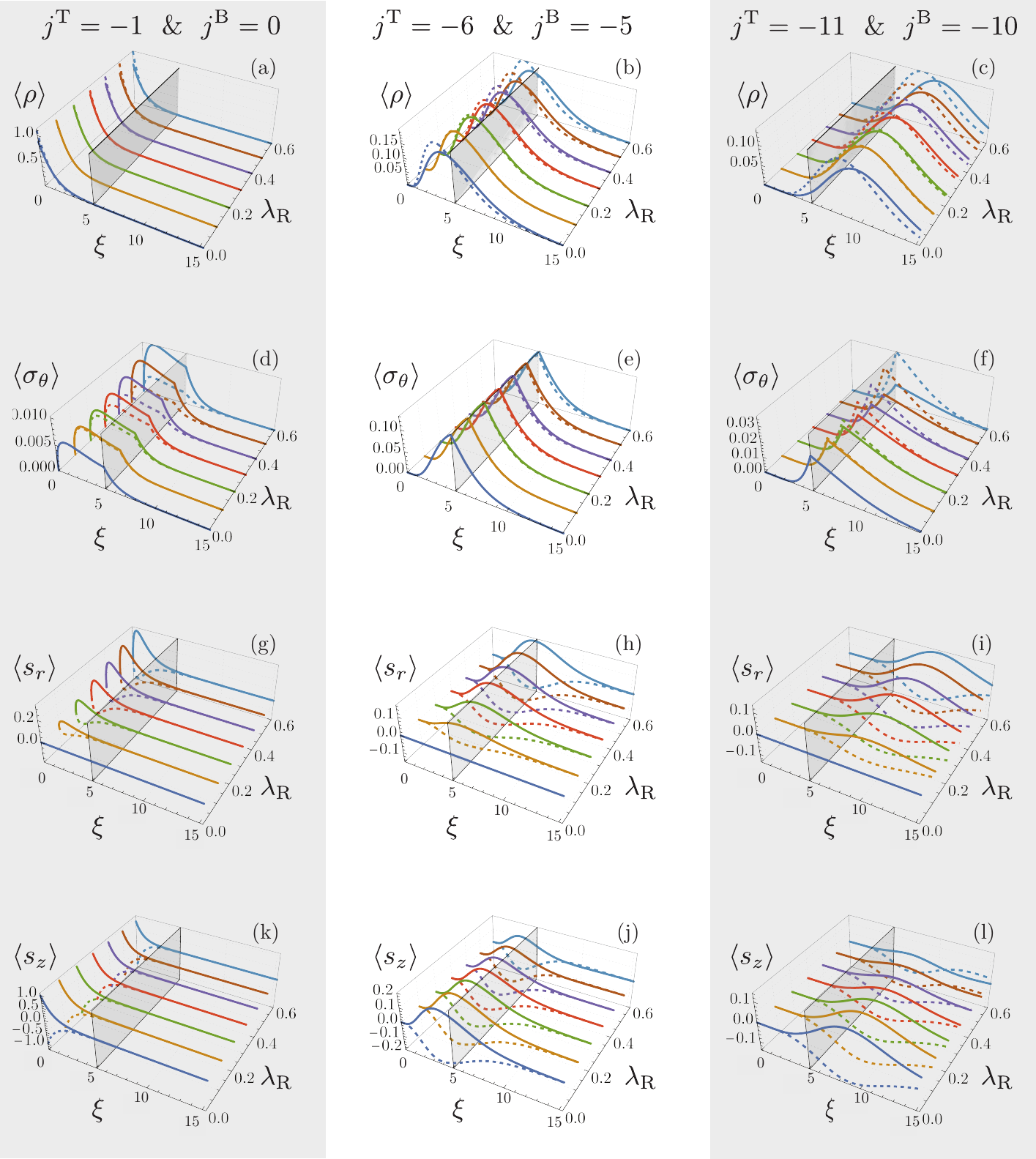}
    \caption{Radial profile of the observable densities defined in Eq.~\eqref{obser}
    for the bottom (solid lines) and top (dashed lines)
    zero modes as a function of $\xi$
    for different values of the Rashba SOC $\lambda_\text{R}$ 
    ($\lambda_\text{R}=0, 0.1, 0.2, 0.3, 0.4,0.5,0.6$).
    Panels (a) to (c) display the modulus square of the wave function  
    for the lowest value of $j$ ($j=-1/0$ for top/bottom mode), 
    an intermediate value of $j$ ($j=-6/-5$ for top/bottom mode), 
    and a large value of $j$ ($j=-11/-10$ for top/bottom mode). 
    Panels (d) to (f), (g) to (i), and (k) to (l) are the same but for
    the densities $\langle\sigma_\theta\rangle$, $\langle s_r \rangle$, and 
    $\langle s_z \rangle$ respectively. 
    In all panels, 
    the plane at $\xi_0=5.1$ represents the position of the $pn$~interface.}
    \label{SM_Fig_4}
\end{figure*}
%
%

In the panels of Fig.~\ref{SM_Fig_4}, we present the radial profiles of various 
observable densities for the top and bottom zero modes for several 
values of the Rashba SOC and for three different values of the angular momentum $j$. 
We denote these quantities as follows:
\begin{equation}
\label{obser}
\langle \rho \rangle = \psi_j^{i\dagger} \psi_j^i,  \quad
\langle \sigma_\theta \rangle = \psi_j^{i\dagger} 
\sigma_\theta \psi_j^{i\dagger}, \quad
\langle s_z \rangle = \psi_j^{i\dagger} s_z \psi_j^i, \quad
\langle s_r \rangle = \psi_j^{i\dagger} s_r \psi_j^i, \quad i=\text{T},\text{B}.
\end{equation}
They describe the radial probability density, the azimuthal probability current density, 
and the perpendicular and radial components of the spin density. 
(See Eqs.~\eqref{pseudospin_operators} and~\eqref{Sigma_theta} 
for the definition of $\sigma_\theta$ and $s_r$.)
Because of symmetry, the two densities $\langle \sigma_r\rangle$ and 
$\langle s_\theta\rangle$ are identically zero.
Panels (a) to (c) in Fig.~\ref{SM_Fig_4} clearly show how the radial profile 
of the probability density changes with the angular momentum $j$: the wave function 
is localized at the center of the circular region for $\{j^\text{T},j^\text{B}\}=\{-1,0\}$, 
it is located close to the $pn$~interface for $\{j^\text{T},j^\text{B}\} =\{-6,-5\}$, and
finally, it moves outside the circular region and away from the $pn$~interface 
for $\{j^\text{T},j^\text{B}\}=\{-11,-10\}$. A similar behavior is found in all 
the other observables we have considered. We note that in the absence of Rashba SOC, 
the radial spin density $\langle s_r\rangle$ vanishes, and it increases 
for increasing values of $\lambda_\text{R}$. On the contrary, the perpendicular spin 
density $\langle s_z\rangle$ is finite also in the absence of Rashba SOC and 
decreases with increasing $\lambda_\text{R}$. 
This can be understood since for increasing $\lambda_\text{R}$ 
the spin will tend to align with the effective magnetic field generated by 
the Rashba SOC~\cite{Frustaglia2004}. 

\comment{This is due to an effective magnetic field experienced 
by the carriers moving along the curved $pn$~junction. The integration 
along the radial direction of these quantities corresponds to the expected value 
that will be used in the next section when we define the effective 1D model 
describing the $pn$~junction.}

\section{Effective zero-mode Hamiltonian}
\label{projection}

In this section, we provide the details of the derivation of the effective 
one-dimensional (1D) Hamiltonian governing the spin and angular dynamics of 
the zero modes localized at the junction. 
We assume that the Fermi energy is at the charge neutrality point, $E_\text{F}=0$, and
we aim at an effective model valid in the low-energy region around $E_\text{F}$.
We follow the approach of~\cite{Meijer_2002}, where 
the effective 1D Hamiltonian for the analogous problem of electrons in a 
mesoscopic ring in the presence of Rashba SOC was derived.
The approach is based on the projection of the full Hamiltonian onto the 
zero-energy radial state, localized at the interface between the $p$ and the $n$ regions. 
This projection is justified as long as the separation between the zero modes 
and the first Landau level is much larger than any other relevant energy scale in the problem. 

We start with the Dirac equation 
$$
\mathcal{H} \Psi =E \Psi,
$$
where the Hamiltonian~\eqref{hamSM} in the symmetric gauge is expressed 
in polar coordinates as follows:
%
%
\begin{equation}
\mathcal{H}  
= \sigma_r (-i\partial_r) + \sigma_\theta \left(\frac{L_z}{r} + A_\theta \right) 
+ V(r) +  \Delta \sigma_z  
+ \lambda_\text{Z}   s_z +  \lambda_\text{KM} \sigma_zs_z
 + \frac{\lambda_\text{R} }{2} (\sigma_r s_\theta-\sigma_\theta s_r),
\end{equation}
%
%
with $A_\theta=r/2$, $V(r)=V_0\, \text{sign}(R-r)$, $L_z=-i\partial_\theta$, 
and we have defined
%
%
\begin{subequations} \label{pseudospin_operators}
\begin{align}
\sigma_r & = \cos\theta \,  \sigma_x + \sin\theta \, \sigma_y =
\begin{pmatrix}
0 & e^{-i\theta} \\
e^{i\theta} & 0 
\end{pmatrix} \label{Sigma_r}, \\ 
\sigma_\theta & = -\sin\theta \, \sigma_x + \cos\theta \, \sigma_y =
\begin{pmatrix}
0 & -i e^{-i\theta} \\
i e^{i\theta} & 0
\end{pmatrix} \label{Sigma_theta}, 
\end{align}
\end{subequations}
%
%
with analogous expressions for $s_r$ and $s_\theta$. First, we make a unitary 
transformation in sublattice and spin space:
%
%
\begin{subequations}
\begin{align}
\tilde{ \mathcal{H}}
& =U \mathcal{H} U^{-1},\\
\tilde \Psi & = U \Psi,
\end{align}
\end{subequations}
%
%
where $U =e^{i \frac{\sigma_z}{2} (\theta + \frac{\pi}{2}) } e^{i \frac{s_z}{2}\theta}$. 
The additional $\pi/2$ rotation in sublattice space is included in order to obtain 
a real Hamiltonian. Using
%
%
\begin{align}
U \sigma_r U^{-1} &= - \sigma_y, \quad U \sigma_\theta U^{-1} =\sigma_x, \label{UnitSigma}  \\
U s_r U^{-1} & = s_x, \quad U s_\theta U^{-1} = s_y,\\
U L_z U^{-1} &= L_z - \frac{\sigma_z}{2} - \frac{s_z}{2},
\end{align}
%
%
we arrive at 
%
%
\begin{equation}
\tilde{\mathcal{H}} 
= i\sigma_y \left( \partial_r +\frac{1}{2r} \right) + \sigma_x \left( \frac{L_z}{r} + \frac{r}{2} - \frac{s_z}{2r} \right) + V(r) 
+ \Delta \sigma_z  +  \lambda_\text{Z}  s_z + \lambda_\text{KM} \sigma_z s_z  
 -  \frac{\lambda_\text{R} }{2}(\sigma_y s_y + \sigma_x s_x).
\end{equation}
%
%
Note that under the unitary transformation $U$, the total 
angular momentum $J$ is mapped to $UJU^{-1}=L_z$. Next, 
we separate the Hamiltonian into a radial part and an angular/spin part, 
$\tilde{\mathcal{H}} = \tilde{\mathcal{H}}_r  + \tilde{\mathcal{H}}_\theta$,
where the radial part is defined as 
%
%
\begin{align}
\tilde{\mathcal{H}}_r =
i\sigma_y \left( \partial_r  + \frac{1}{2r} \right) +
\frac{\sigma_x}{r} \left( \frac{r^2}{2} - \Phi \right)
+V(r)  + \Delta \sigma_z,
\end{align}
%
%
and the angular/spin part as 
%
%
\begin{equation}
\tilde{\mathcal{H}}_\theta
=  \frac{\sigma_x}{r} (L_z +\Phi) +  
\left( \lambda_\text{Z} + \lambda_\text{KM} \sigma_z - \frac{\sigma_x}{2r} \right) s_z  
-  \frac{\lambda_\text{R} }{2} \left( \sigma_y s_y + \sigma_x s_x \right).
\end{equation}
%
%
Here, $\Phi$ is a parameter whose value is set in such a way that the zero mode 
of $\tilde{\mathcal{H}}_r$ is at the Fermi energy $E_\text{F}=0$. 
In practice, $\Phi$ is with good approximation the magnetic flux through the 
$pn$~junction in units of the flux quantum, $\Phi \approx R^2/2\ell_B^2$. 
\comment{If the junction is not sharp, 
there is no clear-cut definition of magnetic flux enclosed by the junction. 
In this case, one can define $\Phi$ as the solution of the equation 
$E_0(\Phi) =E_\text{F}$, where $E_0(\Phi)$ is the spin-degenerate zero-mode 
eigenvalue  of $\tilde{\mathcal{H}}_r(\Phi)$. So, it is the Fermi energy 
that fixes the magnetic flux through the junction.}
The radial Hamiltonian $\tilde{\mathcal{H}}_r$ coincides, up to the $\pi/2$ 
rotation in sublattice space,
with the model of a circular $pn$~junction for spinless graphene solved 
in~\cite{DeMartino_2010,Cohnitz_2016}, 
with the appropriate identification of the parameter $\Phi$. 
\comment{with the spin-down block of $\mathcal{H}_j$ in Eq.~\eqref{ham_bulk} 
with $\lambda_\text{Z}=\lambda_\text{KM}=\lambda_\text{R}=0$ 
and the identification $j=-(R^2+1)/2$. 
}
 
We now project the full Hamiltonian onto the spin-degenerate zero mode 
of $\tilde{\mathcal{H}}_r$. To this aim, we write the wave function as 
%
%
\begin{equation}
    \tilde{\Psi}(r,\theta) = \tilde \psi_0 (r)  \tilde{\chi}(\theta),
\end{equation}
%
%
where the sublattice spinor $\tilde \psi_0(r)$ is the zero mode 
of $\tilde{\mathcal{H}}_r$, which satisfies 
%
%
\begin{equation}
\tilde{\mathcal{H}}_r \, \tilde \psi_0(r) = 0,
\end{equation}
%
%
and we choose to be real, and
$\tilde{\chi}(\theta)$ is a two-component angular spinor in spin space. 
From the equation $\tilde{\mathcal{H}} \tilde{\Psi} = E \tilde{\Psi}$
we find that $\tilde{\chi}(\theta)$ satisfies the equation 
%
%
\begin{equation}
 \tilde{\mathcal{H}}_\text{eff} \tilde{\chi}(\theta) = E \tilde{\chi}(\theta),
\end{equation}
%
%
with the effective 1D Hamiltonian
%
%
\begin{align}
 \tilde{\mathcal{H}}_\text{eff} 
& = \langle \tilde{\mathcal{H}} \rangle_0=
\omega_0  (L_z +\Phi) + \left( \omega_\text{Z} -\frac{\omega_0}{2} \right)s_z   
-  \omega_\text{R} s_x.
\label{effham}
\end{align}
%
%
Here, the brackets $\langle \ldots \rangle_0$ denote the expectation value 
in the radial zero mode:
%
%
$$
\langle \dots \rangle_0 = \int_0^\infty rdr  \, \tilde{\psi}^\dagger_0(r)  
\dots \tilde{\psi}^{}_0(r),
$$
%
%
and we have used $\langle \sigma_y \rangle_0 = 0$, 
which holds because $\tilde{\psi}_0(r)$ is a real spinor. 
In Eq.~\eqref{effham} we have defined the angular velocity $\omega_0$ 
and the Zeeman and Rashba frequencies $\omega_\text{Z}$ and $\omega_\text{R}$, 
as follows:
%
%
\begin{equation}
\label{angular_velocity}
\omega_0 \equiv \left\langle  \frac{\sigma_x}{r} \right\rangle_0 , \quad
\omega_\text{Z} \equiv \lambda_\text{Z}  + \lambda_\text{KM} \langle \sigma_z\rangle_0, \quad
\omega_\text{R}  \equiv \frac{\lambda_\text{R}}{2} \langle \sigma_x \rangle_0 .
\end{equation}
%
%
Since $\sigma_x$ before the unitary transformation was $\sigma_\theta$, see Eq.~\eqref{UnitSigma},
we recognize the coefficient of $L_z$, $\langle \sigma_x/r \rangle_0$, 
as the angular velocity of the circular motion along the junction, 
and the coefficient that renormalizes $\lambda_\text{R}$,
$\langle \sigma_x \rangle_0$, as the azimuthal component of the velocity. Similarly, 
the vanishing of $\langle \sigma_y \rangle_0$ expresses the vanishing of the radial velocity.
We note that if we undo the unitary transformation in spin space, we obtain 
the effective Hamiltonian
%
%
\begin{align}
\mathcal{H}_\text{eff} & 
 = \omega_0  (L_z +\Phi) +  \omega_\text{Z} s_z -  \omega_\text{R} s_r,
\end{align}
%
%
which explicitly shows that the Rashba SOC acts as 
an effective magnetic field that pushes the spin in the in-plane
radial direction. 

It is straightforward to diagonalize $\tilde{\mathcal{H}}_\text{eff}$. 
Its eigenvalues read 
%
%
\begin{equation}
\label{effEig}
E_{0,m,\pm} =  \omega_0  (m+\Phi) \pm  
\sqrt{\left( \omega_\text{Z}-\frac{\omega_0}{2}\right)^2+\omega^2_\text{R}}, 
\end{equation}
%
%
where $m$ is an integer if we impose periodic boundary conditions. 
The corresponding eigenstates are
%
%
\begin{align}
\tilde{\chi}_{m, +}(\theta) =  \frac{e^{im\theta}}{\sqrt{2\pi}}
\begin{pmatrix}
\cos \frac{\gamma}{2} \\
- \sin \frac{\gamma}{2}
\end{pmatrix},\quad 
\tilde{\chi}_{m,-}(\theta)  =   \frac{e^{im\theta}}{\sqrt{2\pi}}
\begin{pmatrix}
\sin \frac{\gamma}{2} \\
\cos \frac{\gamma}{2}
\end{pmatrix} , \label{Effeigenstate}
\end{align}
%
%
where we define
%
%
\begin{align}\label{gamma_angleSM}
\sin \gamma = \frac{\omega_\text{R}}
{\sqrt{ (\omega_\text{Z}-\frac{\omega_0}{2})^2+\omega^2_\text{R} }}, \quad 
\cos \gamma = \frac{\omega_\text{Z}-\omega_0/2}
{\sqrt{ (\omega_\text{Z}-\frac{\omega_0}{2})^2+\omega^2_\text{R} }}.
\end{align}
%
%
Equations~\eqref{effEig} and~\eqref{Effeigenstate} provide 
useful approximation to the SOC coupled zero-mode energies and wave functions, 
which hold as long as transitions to higher Landau 
levels due to $\tilde{\mathcal{H}}_\theta$ can be neglected, 
and for angular states with $m\approx - \Phi$, where it predicts 
a linear dependence of the energy on $m$.

We note that in the uniform system ($V_0=0$), the zero mode of $\tilde{\mathcal{H}}_r$ 
has only one non-vanishing sublattice amplitude (the sublattice pseudo-spin is down-polarized). 
As a consequence, both $\omega_0$ and $\omega_\text{R}$ 
vanish, and the eigenstates are spin-polarized along the $z$ direction 
and orbitally degenerate (i.e., the energy is independent of $m$).
In the presence of the potential step ($V_0 \neq 0$), both sublattice
amplitudes in $\tilde{\psi}_0$ are finite. Then $\omega_0$ and $\omega_\text{R}$ 
are finite, 
the zero modes acquire a dispersion, and the Rashba term is 
activated and
pushes the spin polarization in the planar radial direction.
The spin dynamics is therefore controlled by the potential step amplitude $V_0$.

\section{Rabi condition for general spin-orbit coupling}

\begin{figure}
    \centering
    \includegraphics[width=0.8\textwidth]{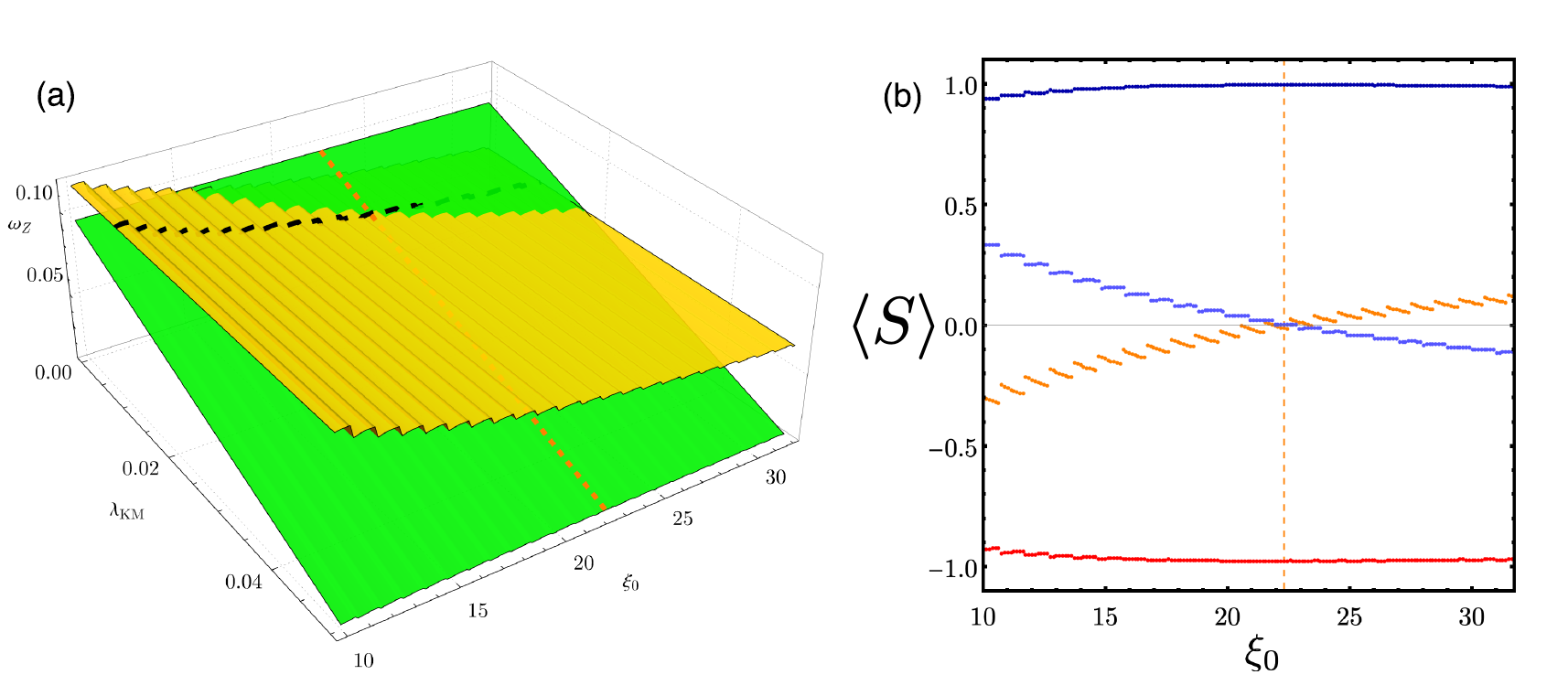}
    \caption{(a) The angular frequency $\omega_0$ in Eq.~\eqref{angular_velocity} 
    (yellow plane) and the effective Zeeman frequency given by Eq.~\eqref{effective_Zeeman} 
    (green plane) as functions of junction size $\xi_0$ and the Kane-Mele SOC $\lambda_\text{KM}$. 
    The intersection of the green and yellow planes corresponds to the sweet spots 
    where the resonance condition occurs.   
    For a given value of $\lambda_\text{KM}$ ($\lambda_\text{KM}=0.01$ in the figure), 
    illustrated by the black dashed line, the value of $\xi_0$ at which the resonance 
    occurs is indicated by the red dashed line. 
    \comment{The sweet spot in the parameter space is found by the intersection of the two planes 
    and the crossing of the dashed black line, corresponding to the value of $\lambda_\text{KM}$.} 
    (b) The exact expectation values of the radial and perpendicular
    spin components in the top and bottom modes as a function of 
    $\xi_0$, for $\lambda_\text{Z}=0.048$ and  $\lambda_\text{R}=0.2$ and $\lambda_\text{KM}=0.01$. 
    The top curve shows $\langle s_r\rangle_\text{B}$, the bottom one $\langle s_r\rangle_\text{T}$, 
    and the two central ones $\langle s_z\rangle_\text{T}$ and $\langle s_z\rangle_\text{B}$. In all the panels, $V_0=0.51$.}
    \label{SM_fig_5}
\end{figure}

In the main text, we have investigated the Rabi condition for the full model 
in Eq.~\eqref{hamSM} under the assumption that only the Rashba SOC is non-vanishing, 
and that effects due to the VZSOC can be neglected | single-valley model. 
We now relax this condition and include the Kane-Mele ($\lambda_\text{KM}$) SOC terms.
In general, the angular frequency associated with the Zeeman term can then be expressed as:
%
%
\begin{equation}\label{effective_Zeeman}  
\omega_\text{Z}=\lambda_\text{Z}+\lambda_\text{KM}\langle\sigma_z\rangle_0.
\end{equation}
%
%
Notice that here the brackets $\langle \dots \rangle_0$ denote the 
expectation value in the $j$-state with energy closest to zero energy at the given value of $\xi_0$.
As already mentioned in the main text, in the single-valley approximation,
the valley-Zeeman SOC just  produces a shift of the Zeeman term. 
The Kane-Mele SOC gives a nontrivial contribution to $\omega_\text{Z}$ that depends 
on the expectation value of $\sigma_z$  over the spinless system. 
In Fig.~\ref{SM_fig_5}, we present the effect of the Kane-Mele SOC 
on the shift of the Rabi condition. From  Fig.~\ref{SM_fig_5}(a), 
we observe for fixed $\lambda_\text{Z}$ that the position 
of the sweet spot $\xi_0$ for increasing values of $\lambda_\text{KM}$ 
moves at larger values of the radius. 
%
%
%
%

\section{$S$-matrix approach} 

\begin{figure*}[!t]
    \centering
    \includegraphics[width=\textwidth]{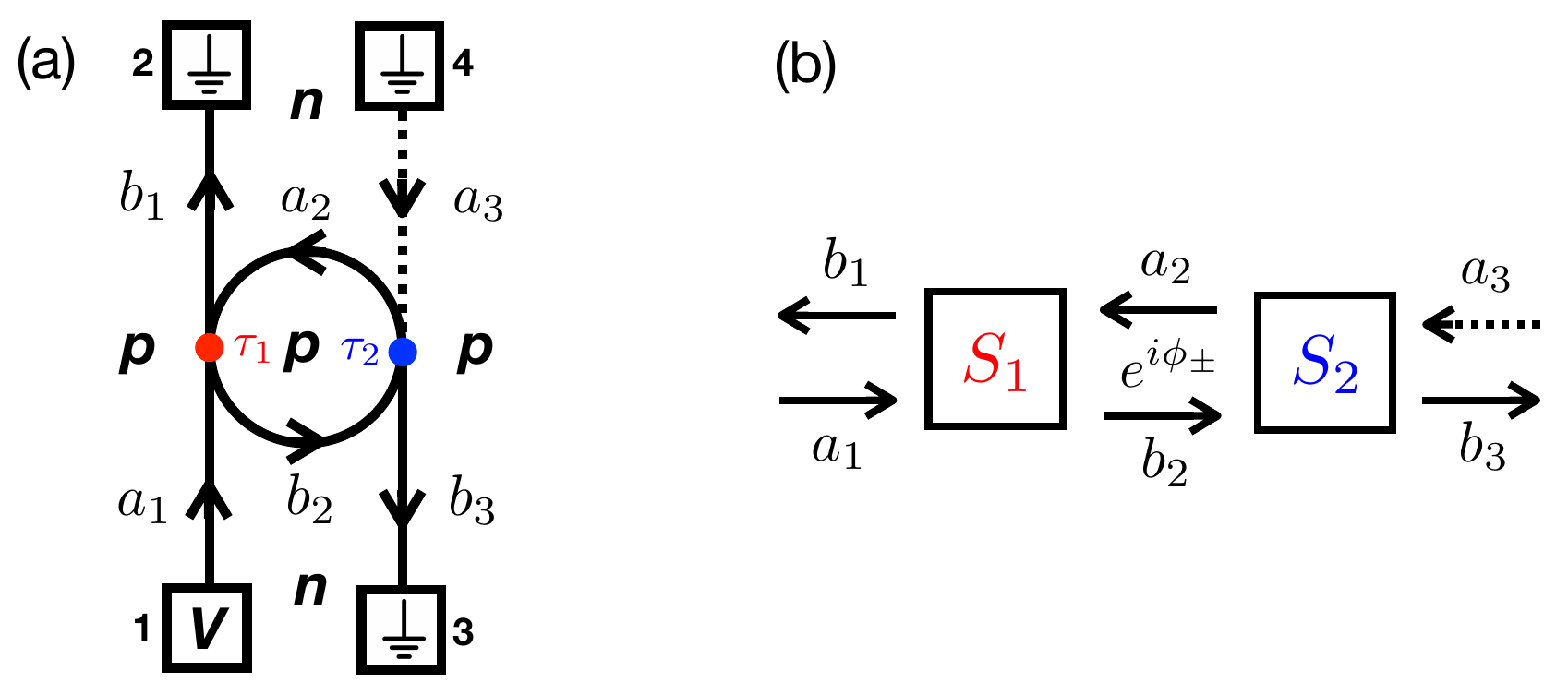}
    \caption{(a) Sketch of the electronic circuit for the case of a $p$-doped circular region. 
    We present the corresponding circuit for the $n$-doped circular region in the main text. 
    (b) Generic representation of two scattering regions $S_1$ and $S_2$ separated by a region 
    where the carriers can gain a geometrical phase $e^{i \phi_\pm}$.}
    \label{fig4}
\end{figure*}

Here we introduce the scattering approach~\cite{Datta1995} used to obtain the quantum transmission 
and conductance of the interferometers discussed in the main text. We begin by discussing 
a spinless model and then generalize it to the spin-dependent case. Without any significant 
loss of generality, we stick to the $ p$-dot-based interferometer depicted in Fig.~\ref{fig4}(a).  

Incoming and outgoing chiral modes are described in Fig.~\ref{fig4} by 
fermionic annihilation operators $\{a_1,a_3\}$ and $\{b_1,b_3\}$, respectively, such that
\begin{equation}
\label{S}
\left( \begin{array}{c}
b_1 \\
b_3
\end{array} \right)=
\left( \begin{array}{cc}
r & t' \\
t & r'
\end{array} \right)
\left( \begin{array}{c}
a_1 \\
a_3
\end{array} \right).
\end{equation}
The conductances $G_{21}$ and $G_{31}$ are determined from the scattering amplitudes $r$ and $t$, 
respectively, by following the Landauer-B\"uttiker approach. 

The scattering matrix on the r.h.s. of Eq.~(\ref{S}) can be obtained by combining 
the scattering blocks $S_1$ and $S_2$ corresponding to the barriers $\tau_1$ and $\tau_2$, 
as depicted in Fig.~\ref{fig4}(b). These block are connected by channels $\{a_2,b_2\}$ 
propagating around the central $p$ dot by accumulating additional phases $\phi_\pm$, satisfying
\begin{eqnarray}
\label{S1}
\left( \begin{array}{c}
b_1 \\
b_2
\end{array} \right)&=&
\left( \begin{array}{cc}
r_1 & t_1' \\
t_1 & r_1'
\end{array} \right)
\left( \begin{array}{c}
a_1 \\
a_2
\end{array} \right),\\
\label{S2}
\left( \begin{array}{c}
a_2 \\
b_3
\end{array} \right)&=&
\left( \begin{array}{cc}
r_2 & t_2' \\
t_2 & r_2'
\end{array} \right)
\left( \begin{array}{c}
b_2 \\
a_3
\end{array} \right),
\end{eqnarray}
with
\begin{eqnarray}
r_1&=&\sqrt{1-\tau_1},~~~ t_1'=e^{i\phi_-}\sqrt{\tau_1},\nonumber\\
t_1&=&e^{i\phi_+}\sqrt{\tau_1},~~~ r_1'=-e^{i(\phi_++\phi_-)}\sqrt{1-\tau_1},\nonumber\\
r_2&=&\sqrt{1-\tau_2},~~~~ t_2'=\sqrt{\tau_2},\nonumber\\
t_2&=&\sqrt{\tau_2},~~~~~~~~~ r_2'=-\sqrt{1-\tau_2}.\nonumber
\end{eqnarray}
After a little algebra, from (\ref{S})-(\ref{S2}) we find
\begin{eqnarray}
\label{r}
r&=&r_1+t_1'(1-r_2r_1')^{-1}r_2t_1,\\
\label{t}
t&=&t_2(1-r_1'r_2)^{-1}t_1.
\end{eqnarray}
Notice that expanding (\ref{r}) and (\ref{t}) as geometric series supplies 
the Feynman paths contributing to the quantum amplitudes due to multiple reflections 
between the barriers. Moreover, when the barriers are placed symmetrically 
on opposite sides 
of the dot we find that $\phi_+=\phi_-$.

The results of Eqs. (\ref{r}) and (\ref{t}) can be generalized to the spin-dependent case 
by choosing convenient spin bases along the linear and circular $pn$~junctions and 
calculating their local projection at barriers 1 and 2. 
For the circular junction, the natural choice is the spin-eigenstate basis, 
which evaluated at the barriers reads
\begin{eqnarray}
\label{spinors2a}
| \chi_+, \ell\rangle &=&
\left( \begin{array}{c}
\cos\frac{\gamma}{2} \\
\ell\sin\frac{\gamma}{2}
\end{array} \right),\\
\label{spinors2b}
| \chi_-, \ell\rangle &=&
\left( \begin{array}{c}
\sin\frac{\gamma}{2} \\
-\ell \cos\frac{\gamma}{2}
\end{array} \right),
\end{eqnarray}
with $\ell=1$ for barrier 1 and $\ell=-1$ for barrier 2.
For linear junctions (acting as incoming and outgoing leads) we can simply choose 
the canonical $z$-basis
\begin{equation}
\label{spinors1}
|\uparrow\rangle =
\left( \begin{array}{c}
1 \\
0
\end{array} \right),\
|\downarrow\rangle =
\left( \begin{array}{c}
1 \\
0
\end{array} \right).
\end{equation}
The use of a field-dependent, spin-eigenstate basis has no practical advantage here 
since the conductance is independent of the spin phases gathered along the leads.
As for the phases $\phi_\pm$, they can be obtained by setting $E_{m,s}=0$ 
in Eq.~\eqref{effEig} and finding the corresponding spin-dependent $m$ (which 
is not necessarily an integer any longer due to the open boundary conditions 
introduced by the barriers). As a result, we find
\begin{equation}
\label{phases}
    \phi^s=-s\pi\sqrt{\left(Q_{\rm Z}-\frac{1}{2}\right)^2+Q_{\rm R}^2}-\pi\Phi, \quad s=\pm,
\end{equation}
with $Q_{\rm R}=\omega_{\rm R}/\omega_0$, $Q_{\rm Z}=\omega_{\rm Z}/\omega_0$, and 
where we have dropped the $\pm$ subindex due to symmetry.

The spin-dependent scattering amplitudes are determined  by the projections 
$\langle \uparrow|\chi_s,\ell\rangle$ and $\langle \downarrow|\chi_s,\ell\rangle$, 
and by the phases (\ref{phases}), accordingly. In this way, we find
\begin{eqnarray}
\label{r1}
r_1&=&\sqrt{1-\tau_1}
\left( \begin{array}{cc}
1 & 0 \\
0 & 1
\end{array} \right),\\
\label{t1}
t_1&=&\sqrt{\tau_1}
\left( \begin{array}{cc}
e^{i\phi^+} \cos\frac{\gamma}{2} & -e^{i\phi^+} \sin\frac{\gamma}{2} \\
e^{i\phi^-} \sin\frac{\gamma}{2} & e^{i\phi^-} \cos\frac{\gamma}{2}
\end{array} \right),\\
\label{t1p}
t_1'&=&\sqrt{\tau_1}
\left( \begin{array}{cc}
e^{i\phi^+} \cos\frac{\gamma}{2} & e^{i\phi^-} \sin\frac{\gamma}{2} \\
-e^{i\phi^+} \sin\frac{\gamma}{2} & e^{i\phi^-} \cos\frac{\gamma}{2}
\end{array} \right),\\
\label{r1p}
r_1'&=&-\sqrt{1-\tau_1}
\left( \begin{array}{cc}
e^{i2\phi^+} & 0 \\
0 & e^{i2\phi^-}
\end{array} \right),\\
\label{t2}
t_2&=&\sqrt{\tau_2}
\left( \begin{array}{cc}
\cos\frac{\gamma}{2} & \sin\frac{\gamma}{2} \\
\sin\frac{\gamma}{2} & -\cos\frac{\gamma}{2}
\end{array} \right),\\
\label{r2}
r_2&=&\sqrt{1-\tau_2}
\left( \begin{array}{cc}
1 & 0 \\
0 & 1
\end{array} \right).
\end{eqnarray}
Here we omit the spin-dependent expressions for $t_2'$ and $r_2'$ 
since they do not contribute to the scattering amplitude matrices $t$ and $r$ in 
Eqs. (\ref{r}) and (\ref{t}). By following the Landauer-B\"uttiker approach, 
we find the expression for the linear conductances in terms of $t$ and $r$ that
we use in the main text:  
\begin{eqnarray}
\label{G21}
G_{21}&=&\frac{e^2}{h}\tr[rr^\dagger],\\
\label{G31}
G_{31}&=&\frac{e^2}{h}\tr[tt^\dagger].
\end{eqnarray}
%
%
%
%
%

\section{First-order expansion} 
\label{1stOE}

The scattering amplitudes (\ref{r}) and (\ref{t}) can be expanded as 
\begin{eqnarray}
\label{rexp}
r&=&r_1+t_1'r_2t_1+t_1'r_2r_1'r_2t_1+...,\\
\label{texp}
t&=&t_2t_1+t_2r_1'r_2t_1+...
\end{eqnarray}
These expansions have a simple interpretation in terms of Feynman paths: 
each term corresponds to a possible scattering history through barriers 1 and 2, 
comprising a different number of windings around the central dot. Previous works 
in semiconductor-based rings~\cite{Frustaglia2004,Nagasawa2013} have shown that 
the first two terms in the expansions (\ref{rexp}) and (\ref{texp}), corresponding 
to zero- and single-winding path contributions, are sufficient to capture all 
relevant features of the conductances (\ref{G21}) and (\ref{G31}). This also 
facilitates further physical insight by discriminating the role that different 
quantum phases play in the interference. By setting $\tau_1=\tau_2=1/2$ 
and neglecting higher order contributions, we find
\begin{equation}
\label{G211stSM}
G_{21} \approx 1+ \cos\phi_\text{AB} \cos\phi_\text{S},
\end{equation}
with 
\begin{eqnarray}
\label{ABphaseSM}
\phi_\text{AB}&=& 2\pi \Phi,\\
\label{SphaseSM}
\phi_\text{S}&=& 2\pi \sqrt{\left(Q_{\rm Z}-\frac{1}{2}\right)^2+Q_{\rm R}^2},
\end{eqnarray}
where $\phi_\text{AB}$ and $\phi_\text{S}$ are independent phase contributions with origin in the orbital and spin degrees of freedom, respectively.

\section{Effects of valley-Zeeman coupling}
%
%
\begin{figure*}[!t]
    \centering
    \includegraphics[width=0.5\textwidth]{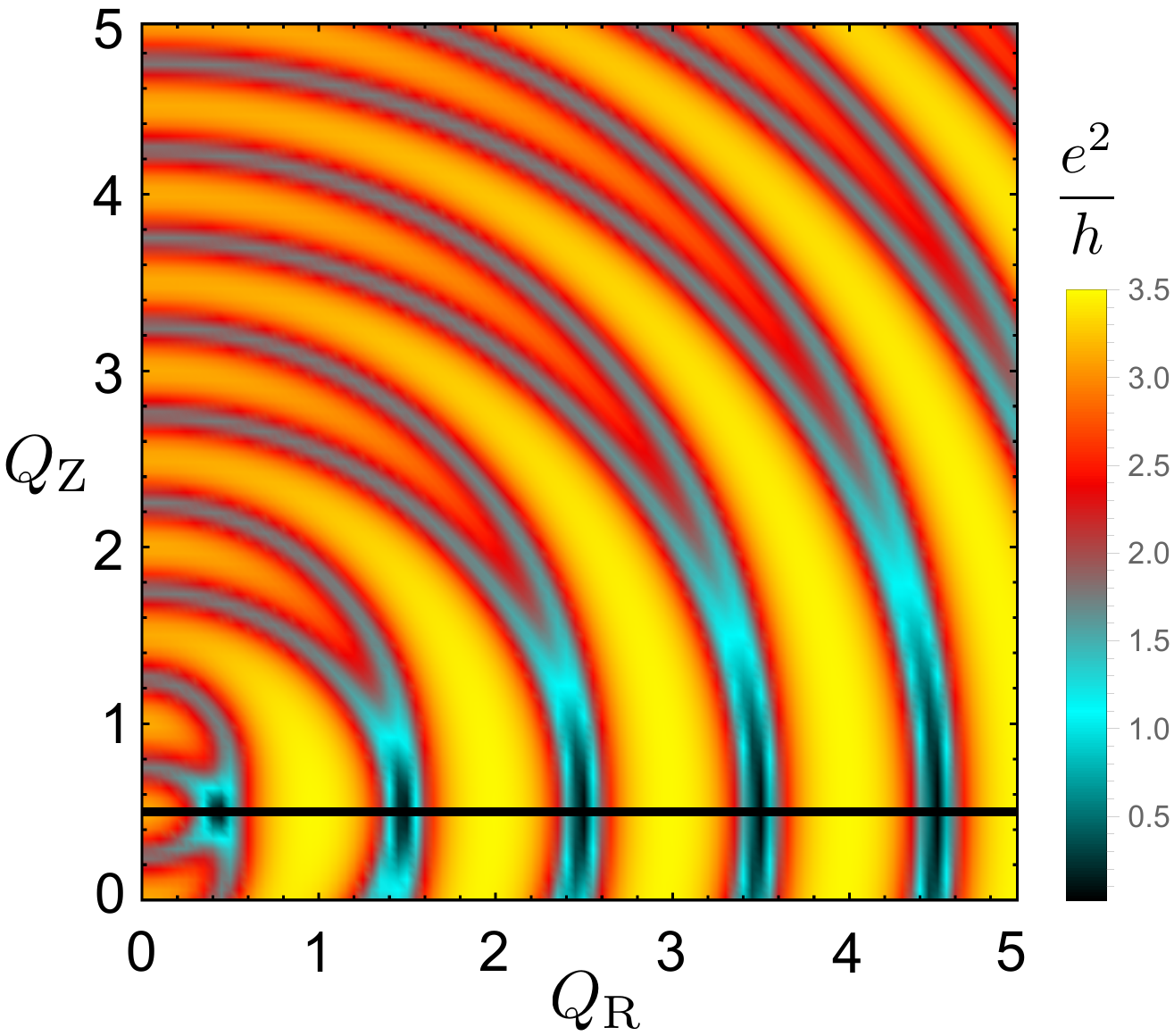}
    \caption{Differential conductance $G_{21}=G_{21}^++G_{21}^-$ for the circuit of 
    Fig. 3(b) including a valley-Zeeman 
    $Q_{\rm VZ}=0.25$. The solid line at $Q_{\rm Z}=1/2$ corresponds to a symmetry axis in the pattern.}
    \label{fig7}
\end{figure*}
%
%
The introduction of the valley-Zeeman coupling leads to a valley-dependent shift in the Zeeman term (a valley splitting), 
such that $\lambda_\text{Z} \rightarrow \lambda_\text{Z} +\tau \lambda_\text{VZ}$, with $\tau=\pm 1$ the valley index. 
This implies that the sweet spots corresponding to in-plane spin eigenstates and $\pi$ geometric phases shift as well,
in a valley-dependent way.
By assuming no valley mixing (due to the separation between lattice and $pn$-junction length scales), 
the conductance turns valley-dependent. The 1st order expansion of Sec. \ref{1stOE} generalizes to
%
%
\begin{equation}
\label{G211stVZ}
G_{21}^\tau \approx 1+ \cos\phi_\text{AB} \cos\phi_\text{S}^\tau,
\end{equation}
%
%
with 
%
%
\begin{equation}
\label{SphaseVZ}
\phi_\text{S}^\tau= 2\pi \sqrt{\left(Q_{\rm Z}+\tau Q_{\rm VZ}-\frac{1}{2}\right)^2+Q_{\rm R}^2},
\end{equation}
%
%
and $Q_{\rm VZ}=\lambda_{\rm VZ}/\omega_0$. The valley-resolved conductances for circuits with $n-$ 
and $p-$doped central regions would then look like those of Figs. 3(c) and 3(d) with additional 
$\pm Q_{\rm VZ}$ shifts along the $Q_{\rm Z}$ axis, respectively. 
As for the total conductance, it is the sum of the corresponding valley conductances, 
$G_{21}=G_{21}^++G_{21}^-$. In Fig. \ref{fig7} we illustrate this situation 
for the $p$-doped dot circuit of Fig. 3(b), with the same setting used to produce 
Fig. 3(d) and an additional valley-Zeeman coupling $Q_{\rm VZ}=0.25$. 
The composed pattern is symmetric with respect to the axis $Q_{\rm Z}=1/2$. 
Although this axis does not correspond any longer to a sweet spot in the strict sense discussed above,  
we notice that the average spin projection along $z$ vanishes at $Q_{\rm Z}=1/2$, 
i.e., $\langle s_z\rangle=\langle s_z\rangle_++\langle s_z\rangle_-=0$, due to the opposite valley-Zeeman pulls.

\end{document}